\documentclass[reprint,amsmath,amssymb,aps,prx]{revtex4-2}

\usepackage{graphicx}
\usepackage{dcolumn}
\usepackage{bm}

\usepackage{physics}
\usepackage{booktabs} 
\usepackage{exscale}
\usepackage{xcolor}
\usepackage{comment}
\usepackage{wrapfig}
\usepackage{lipsum}
\usepackage{mathtools}
\usepackage{multirow}
\usepackage{tikz}
\usetikzlibrary{decorations.pathreplacing,calligraphy}
\usepackage{cancel}

\usepackage[normalem]{ulem}

\usepackage{cancel}

\newcommand				{\cumulant}[1]	{ \expval{\!\!\expval{#1}\!\!}}

\renewcommand			{\v}			{\hat{v}}

\newcommand				{\p}			{\hat{p}}

\newcommand				{\X}			{\hat{X}}
\renewcommand			{\P}			{\hat{P}}
\newcommand				{\x}			{\hat{x}}
\newcommand				{\n}			{\hat{n}}
\newcommand				{\m}			{\hat{m}}
\newcommand				{\dn}			{\delta \hat{n}}
\newcommand				{\N}			{\hat{N}}

\renewcommand			{\a}			{\hat{a}}
\renewcommand			{\b}			{\hat{b}}
\newcommand				{\bdag}[1][]    {\hat{b}^{\dag #1}}

\newcommand				{\adag}[1][]    {\hat{a}^{\dag #1}}

\newcommand				{\Te}			{T_\text{e}}
\newcommand				{\Vdc}			{V_\text{dc}}

\newcommand				{\GHz}			{\text{$\mathrm{GHz}$}}
\newcommand				{\MHz}			{\text{$\mathrm{MHz}$}}

\newcommand				{\Hz}			{\text{$\mathrm{Hz}$}}

\newcommand				{\Ohm}			{\text{$\mathrm{\Omega}$}}

\newcommand				{\mK}			{\text{$\mathrm{mK}$}}

\newcommand				{\dB}			{\text{$\mathrm{dB}$}}

\newcommand				{\GSaPerSec}	{\text{GSa/sec}}

\newcommand				{\hc}			{\text{h.c.}}

\DeclareMathOperator	{\conv}			{\circledast}

\hfuzz=\maxdimen \tolerance=10000
\begin{document}

\preprint{APS/123-QED}

\title{Counting statistics of ultra-broadband microwave photons} 

\author{Simon Bolduc Beaudoin}
\author{Edouard Pinsolle}
\author{Bertrand Reulet}
\affiliation{Département de physique, Institut quantique, Université de Sherbrooke, Sherbrooke, Québec, Canada J1K 2R1
}

\date{\today}

\begin{abstract}
We report measurements of counting statistics, average and variance, of microwave photons of ill-defined frequency : bichromatic photons, i.e. photons involving two well separated frequencies, and "white" broadband photons. Our setup allows for the analysis of single photonic modes
of arbitrary waveform over the 1-10 GHz frequency range.
The photon statistics are obtained by on-the-fly numerical calculation from the sampled time-dependent voltage. 
Using an ac+dc biased tunnel junction as a source of quantum microwave, we report an ultra-wide squeezing spectrum representing a competitive source for entanglement generation (up to 0.9 billion measured entangled bits per second) easily achievable experimentally.
We also report the observation of quantum steering by the tunnel junction, and show how the presence of squeezing of a broadband mode implies the existence of entanglement between two modes it encompasses. 
 
\end{abstract}
\maketitle

Quantum correlations in the electromagnetic field is a key resource in the development of quantum technologies.
Entangled light can substantially improve the signal to noise (SNR) ratio in imaging \cite{Lloyd_2008,Barzanjeh_2015}, gravitational wave detection \cite{Ligo_2011} and is currently used to accelerate dark matter research \cite{Backes2021}.

To achieve high entanglement generation rates in the microwave domain, a great effort has been put into adapting traveling wave parametric amplifiers (TWPA) \cite{esposito_TWPA2022,Qiu_2023,Hakonen_2022} to produce broadband squeezed states. 
Widening the band over which squeezing occurs increases the number of useful entangled pairs in the signal.
Indeed, TWPAs are outperforming Josephson parametric amplifiers (JPAs) by their squeezing bandwidth, from 0.7 \GHz \cite{Roy2015a,Mutus2014} to 4 \GHz \cite{Hakonen_2022}, while also promising a large amount of squeezing. 
While TWPAs are a fascinating and promising technology for the generation of large bandwidth quantum states, they are complex to design and very difficult to fabricate.

Developing broadband sources also requires developing the theoretical and experimental tools to analyze their output field. 
A usual characterization consists of detecting the quadratures of the field at pairs of frequencies $F-\Delta f$ and $F+\Delta f$ symmetrically located around $F$. The central frequency $F$ is related to the frequency $f_p$ of the pump used to generate pairs in the TWPA: $F=f_p/2$ for three-wave mixing, $F=f_p$ for four-wave mixing. 
The so-called squeezing spectrum, which measures the amount of squeezing vs. $\Delta f$ 
 is usually obtained by correlating the quadratures at frequencies $F+\Delta f$ and $F-\Delta f$. Detecting these quadratures requires knowing the phase of the pump, for example to phase-lock a local oscillator used to down-convert the measured signal. But one may want to know whether photon pairs are present in the signal irrespective of any phase reference. 
This is the case, for example, when the DC Josephson effect is used to generate correlations \cite{Jebari_2018, Rolland_2019, Peugeot_2021, Menard_2022, Albert_2024}, or in the more general situation where one does not know the exact conditions generating the signal of interest as for the study of correlations in the cosmic microwave background \cite{CMB_2021}. 

In this article we develop another way of analyzing broadband signals, by measuring the counting statistics of single photonic modes that are not monochromatic but broadband, i.e. with no well-defined frequency. 
We achieve this goal by numerical treatment of the measured time-dependent voltage, following the theory developed in \cite{Virally2018}. 
We focus on two photonic modes: i) bichromatic photons, i.e. photons of two simultaneous frequencies, and ii) ”white” photons,i.e. single modes having a broad frequency content, up
to 1 to 10 GHz.

Selecting numerically the mode basis is highly flexible and could be particularly useful for coherent wideband emitters such as pulses of sub-cycle duration \cite{Krauss_2010,Wirth_2011,Riek_2015,Onoe_Husimi}.
We chose to work with a normal-metal tunnel junction in the microwave regime (1 to 10 GHz) as a simple wideband emitter to test this method. This device is easy to fabricate, yet provides a quantum radiation with a bandwidth only limited by the RC time of the junction, here $>15$ GHz. Indeed, in the presence of a dc+ac bias, such a junction has been predicted to emit broadband squeezed radiation \cite{bbrb2013}, which has been confirmed experimentally on various narrow bands \cite{gasse_observation_2013,forgues_non-classical_2016}. While using bichromatic modes our technique provides the whole squeezing spectrum of the junction, working with wideband modes allows us to observe squeezing with larger bandwidth and improved signal to noise ratio.

Our experiment also allows for the measurement of quantum information related correlations for broadband signals: entanglement of formation \cite{Adesso_2005}, entanglement generation rates \cite{Hakonen_2022,esposito_TWPA2022} and steering \cite{Kogias_2015,Wen_2020}. 
Such quantities are essential in quantum cryptography or quantum computing. For example, in applications such as  continuous variable quantum key distribution (CV-QKD), one may want to maximize discord \cite{Su_2014}, entanglement \cite{Adesso_2005} or steering \cite{Kogias_2015} at the highest possible rate. 
We indeed observe an entanglement generation rate comparable to TWPAs: the weak amount of squeezing is compensated by the large bandwidth over which photon pairs are generated. 
We also show that the high frequency part  (6-9 GHz) of the spectrum of the radiation of the tunnel junction could steer the low frequency part of it (3-6 GHz).

\section{Principle of the measurement}

We consider experiments in the microwave domain where a source placed at ultra-low temperature generates an electromagnetic radiation that propagates along a coaxial cable and is detected by a matched amplifier. 
In this setup the measured quantity is the time-dependent voltage amplitude  of that wave, represented by the operator $\v$, superimposed on the noise of the amplification scheme. 
After further amplification, the signal is digitized at high speed and all the relevant quantities are computed from the digitized signal.

\subsection{First quantization: definition of modes}
In order to compute the photon statistics of the signal, one has to define which photons to count. In \cite{Virally2018} the chosen basis was that of photons localized in time, at the extreme opposite of the usual first quantization in the frequency domain. 
In a real experiment, one has to mitigate both approaches by considering wavelets of finite time and frequency spread. 
We note $\beta(t)$ such wavelets in time domain and $\beta(f)$ in frequency domain; they correspond to the current mode of interest. Since $\beta(t)$ is real, $\beta(-f)=\beta^*(f)$. 
We start by considering two such wavelets $\beta_{1,2}$ centered around frequencies $f_{1,2}$ that are well separated so that $\beta_1$ and $\beta_2$ do not overlap. 
Typically, the width of $\beta_{1,2}(f)$ is $\sim200$MHz whereas $f_2-f_1$ varies between 1 and 10 GHz. We define the annihilation operators $a_{1,2}$ of photons in the modes $\beta_{1,2}$ by:
\begin{equation}
    \a_{1,2}=\int_0^{+\infty}\beta_{1,2}(f)\a(f)df \;\;\; ,
\end{equation} where $\a(f)$ is the usual annihilation operator for a photon at frequency $f$. 
Since the wavelets $\beta_{1,2}(f)$ are narrow in frequency domain, one can simply think of the operators $\a_{1,2}$ as annihilating photons of frequency $f_{1,2}$ which would correspond to the limit $\beta_{1,2}(f)=\delta(f-f_{1,2})$.  The $\beta$ functions are normalized according to:
\begin{equation}
    \int_{0}^{+\infty} |\beta(f)|^2 \dd{f}=1 \;\;\; .
\end{equation}
since they correspond to single photonic modes (see e.g. \cite{barnett_methods_2002}). From this we define the annihilation operator for a bichromatic photon,
\begin{equation}
    \b=\frac{\a_1+\a_2}{\sqrt2} \;\;\; .
\end{equation}

This operator annihilates a photon of mode $\beta=(\beta_1+\beta_2)/\sqrt2$. 
Here the plus sign between the monochromatic annihilation operators $\a_{1,2}$ is arbitrary. One may consider a minus sign or any complex phase as well (we studied experimentally this for the photon statistics of squeezed radiation, see below). The photon number operator is defined by:
\begin{equation}
    \n=\bdag\b=\frac{ \n_1 + \n_2}{2} + \hat{\Delta} \;\; , \;\; \text{with } \hat{\Delta} = \frac{\adag_1\a_2 +\adag_2\a_1}{2} \; . 
    \label{eq:Delta}
\end{equation}

The photocount distribution of bichromatic photons is obtained by calculating the moments of $\n$.
$\n$ is distinct from the total photon number operator $\N = \adag_1\a_1 + \adag_2\a_2$ (see e.g. \citet{barnett_methods_2002} chapter 3). 
The former is a measure of the photon number in a single particular mode of the field as the latter is summed over all modes. Here for example, we can complete the 2-frequency basis by introducing $\hat c = (\a_1-\a_2)/\sqrt2$ such that $\N = \bdag\b+\hat c^\dag\hat c$, $[\bdag,\b] = 1$ and $[\bdag,\hat{c}] = 0$.

\begin{figure}[!ht]
\includegraphics[width=\columnwidth]{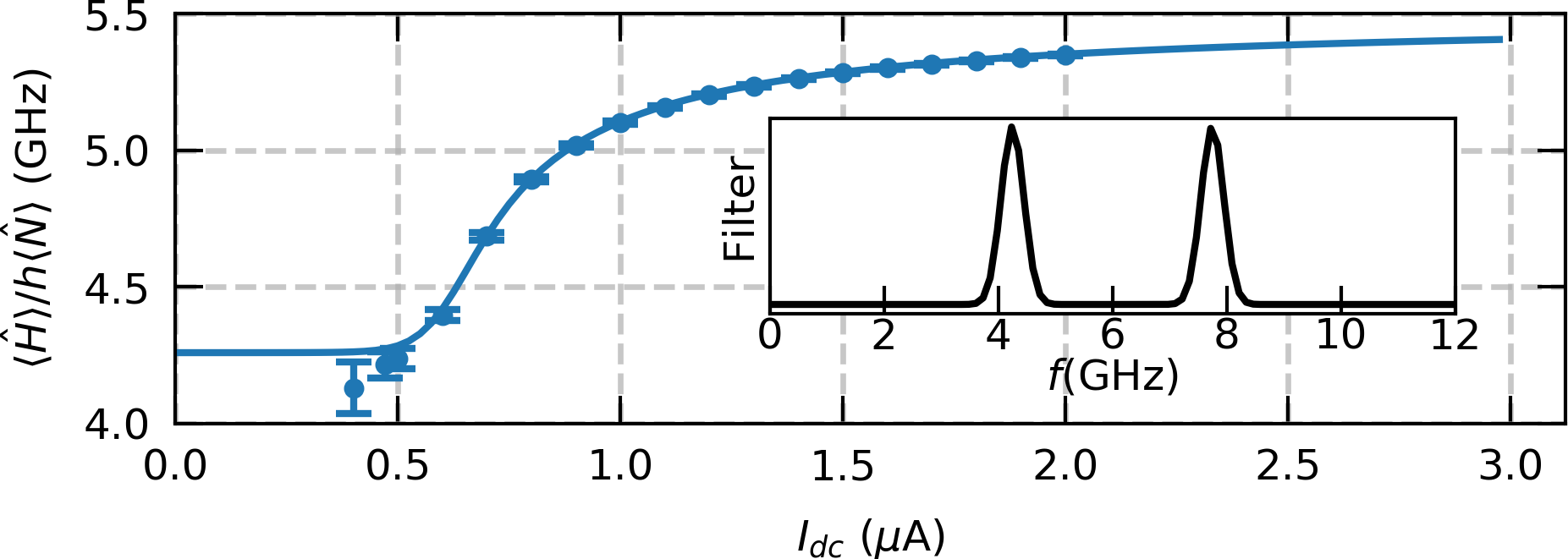}
\caption{ Total average energy $\langle\hat H\rangle$ divided by the total average photon number $\langle \hat N\rangle$, in units of frequency in GHz for bichromatic photons (at frequencies $f_1=4.25~\GHz$ and $f_2=7.75~\GHz$) emitted by the tunnel junction as a function of the dc current bias. Markers are experimental data, solid line theoretical prediction. There is no choice of frequency $\bar{f}$ for which $\langle \hat H\rangle=\langle \N\rangle h\bar{f}$. Data at very low current were omitted since both $\langle \hat N\rangle$ and $\langle \hat H\rangle$ vanish in this limit. 
Inset: The numerical filter used for this experiment.}
\label{fig:NandHvsIdc}
\end{figure}

Since we are dealing with broadband signals, the total number of photons $\hat{N}$ and the total energy $\hat{H}=\hbar\omega_1\n_1+\hbar\omega_2\n_2$ are not proportional. 
We illustrate this point experimentally in Fig.\ref{fig:NandHvsIdc}: we measure, on the one hand, the total energy $\hat H$ and on another the total photon number $\hat N$ of the radiation of a tunnel junction in modes of well separated frequencies $f_1=4.25~\GHz$ and $f_2=7.75~\GHz$.
We plot the "effective frequency" $\expval{\hat H}/\expval{h\hat N}$ as a function of the current bias in the junction. 
It is clear that it is not a constant: it varies from $\sim4.2$ GHz at low bias to $\sim5.5$ GHz at high bias, as a direct result of the radiation being broadband.

\subsection{Quadrature Transform}
The link between the measured time-dependent voltage $\hat{v}(t)$ and the photocount statistics is performed by defining the quadrature $\hat{x}(t)$ according to \cite{Virally2018} :
\begin{equation}
		\x(t) =\beta\conv k\conv\v\,(t)\;, 
  \label{eq:convolution}
\end{equation}
where $\conv$ is the convolution product and $k(t)=\sqrt{1/(Zh|t|)}$ with $Z$ the impedance of the detector, across which $\v$ is measured. $\x$ is the generalization to broadband signals of the usual quadrature of a signal at a given frequency $f$, i.e. its in-phase component. In frequency space, $\x(f)=\beta(f) k(f)\v(f)$ with
\begin{equation}
\hat{v}(f)=-i\sqrt{\frac{Zhf}{2}}\a(f)e^{-2i\pi ft} + \mathrm{H.c.}  \;\; .  
\label{eq:v}
\end{equation}
The kernel $k$ is chosen such that $\x(f)=[\a(f)+\adag(f)]/\sqrt2$
and $\beta$ defines the photonic mode, as discussed above. More precisely, the kernel $k(t)$ transforms the measured voltage $\hat v(t)$ into the quadrature $\hat x(t)$ by canceling out the frequency-dependent prefactors which relate $\hat v$ to $\hat a$ in Eq.(\ref{eq:v}).

\subsection{Photocount statistics}
From the computed time-dependent quadrature, we can deduce the photocount statistics. 
Herein we focus on the average photon number , $\expval{\n}$, and its variance , $\expval{\dn^2}$, with $\dn=\n-\expval{\n}$. 
The procedure is similar to what has been done in \cite{virally_discrete_2016,simoneau_photon_2017}, although differences arise from the large bandwidth involved here (see appendix \ref{Annexe:Quad_mesure}).
We find:
\begin{eqnarray}
        \expval{\n}       &=& \expval{\x^2} - \frac{1}{2}
    \nonumber\\
        \expval{\dn^2}    &=& \frac{2}{3} \expval{\x^4} - \expval{\x^2}^2 - \frac{1}{4}
        ~~~,~
\end{eqnarray}
where the average over $\x^2(t)$ and $\x^4(t)$ is experimentally taken over time. 
This is a generalization of the theoretical results of \cite{virally_discrete_2016,simoneau_photon_2017} for any mode $\beta$ with certain restrictions concerning non-symmetric (n.s.) terms (i.e. terms that are non-symmetric with respect to conjugation, ex: $\bdag\b^3$). These terms are set to be negligible either by construction (choosing $\beta(f)$ such that they are guaranteed to be 0) or by supposing that the light source guarantees $\expval{\text{n.s.}} = 0$, which is the case for a tunnel junction.

\section{Methodology}
\label{Sec:Metho}

\subsection{Experimental setup}
\begin{figure}[!ht]
\includegraphics[width=\linewidth]{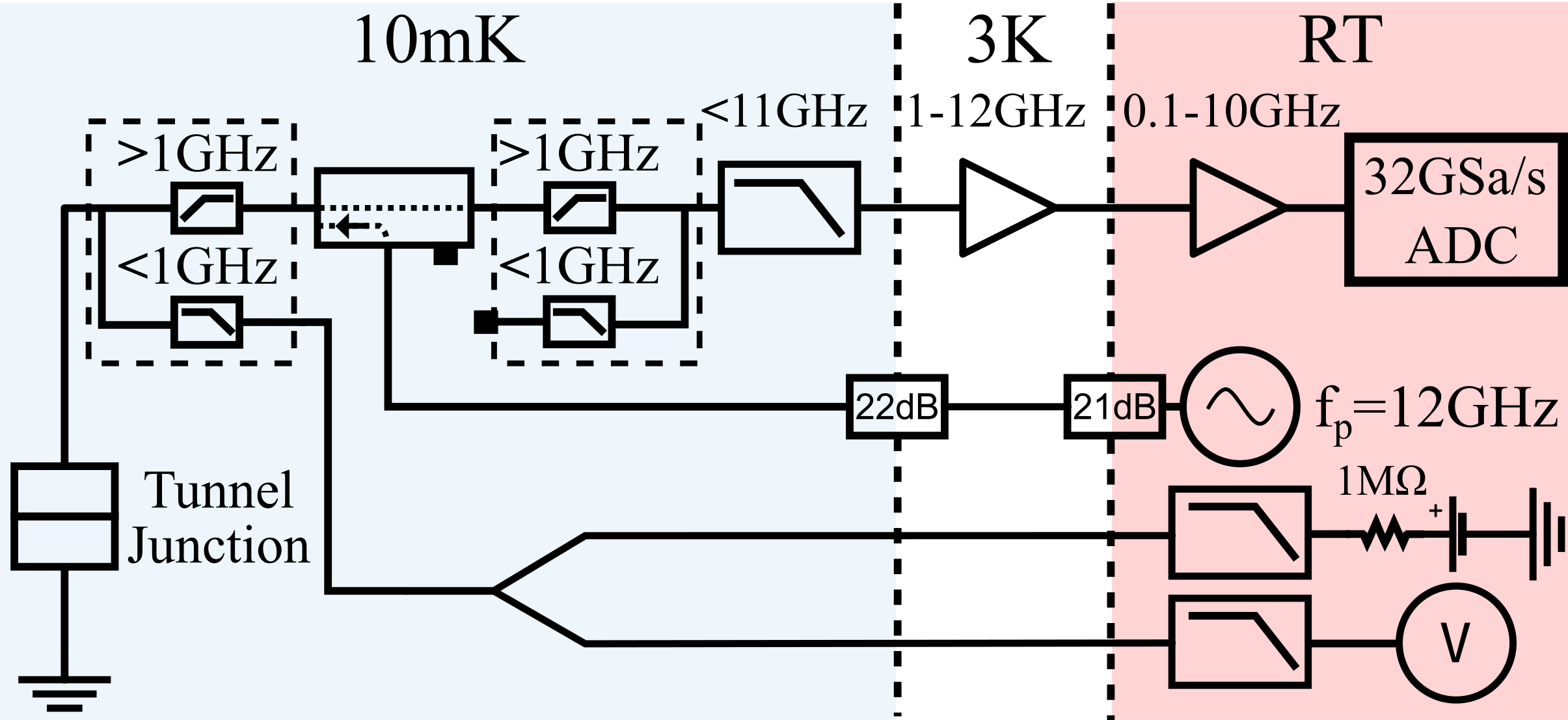}
\caption{ Experimental setup.  On the top is the detection branch, with a bandwidth from 1 to 10 GHz. The middle branch is that of ac bias, with 79 dB of attenuation (43 dB attenuator + 20 dB coupler + 16 dB cables). The lowest branch is the dc line, with a current source (a voltage source in series with a 1M$\Omega$ resistor) and a voltmeter to measure the dc voltage across the junction. Both are filtered by a low-pass (6 Hz) filter at room temperature (RT) while thermocoax between RT and the dilution stage strongly attenuate noise above 10MHz.
Colors represent the temperature of the various stages: blue is 10 mK, white is 3 K and red is RT.
}\label{fig:setup}
\end{figure}

The experimental setup is depicted in Fig. \ref{fig:setup}. 
The source of the electromagnetic field is an Al / $\mathrm{AlO_x}$  / Al tunnel junction 
of resistance $R=52.5~\Ohm$ fabricated using standard photolithography and e-beam evaporation techniques \cite{Spietz_2003}. 
It is placed on the cold stage of a dilution refrigerator with a base temperature of $7$ mK. The superconductivity in aluminum is suppressed by the stray magnetic field $\sim 100$ mT of a strong rare earth magnet placed underneath the sample holder. In applications where a magnetic field in undesirable, the superconductivity of the junction could be suppressed e.g. by adding a small amount of magnetic impurities to the aluminum electrodes \cite{Ruggiero_2004}.

The microwave radiation generated by the junction comes out of the high frequency ($>1~\GHz$) branch of a diplexer pair, then is low-pass filtered $(<11~\GHz)$, goes through a cryogenic 1-12 GHz HEMT (LNF-LNC1-12A) amplifier followed by a room temperature amplifier and is finally digitized using a 10 bit, $32~\GSaPerSec$ digitizer with $10~\GHz$ analog bandwidth (Guzik ADP7104), resulting in a $1.25-10.25~\GHz$ total analog bandwidth.

The ac excitation reaches the sample through attenuated stainless steel coaxial cables followed by the $-20~\dB$ port of a directional coupler. The excitation frequency $f_p=12~\GHz$ is above the cutoff of the low-pass filter inserted in the detection line (-17.5 dB insertion loss at 12 \GHz), so its weak (-16dB) reflection on the junction does not affect the measurement by e.g. inducing nonlinearities.

The dc ports are low-pass filtered ($<6~\Hz$) at room temperature before reaching the sample through Thermocoax cables ($<1~\MHz$) all the way between room temperature and the $7~\mK$ stage.

\paragraph{ Achieving  very low temperature without circulators.}
Special care was crucial to reach an extremely low electron temperature $\Te <20~\mK$. A first diplexer in front of the amplifier removes the low frequency noise (0-1 GHz) generated by the amplifier. A second diplexer connected to the sample plays the role of a bias tee without adding noise. Caution has been taken to avoid ground loops, in particular those involving the cryogenic amplifier's power supply and the communication channels (USB, etc.).

\paragraph{ Calibration  }
The power gain and electronic temperature, $\Te$, are continuously measured and fitted from the full spectrum of current auto-correlations, as in \cite{Spietz_2003,forgues_non-classical_2016,simoneau_mesures_2021,esposito_TWPA2022}. 
The only critical calibration parameter needed to recover photon statistics is the gain (between the sample and the A/D converter) as it is used in the kernel for numerical deconvolutions (see Eq.\eqref{eq:time_quad_numerical} and appendix \ref{Annexe:Gain}). 
The electronic temperature and resistance of the junction, as well as and the loss on the ac port at 12 GHz, are necessary to compare our measurements with theory.
The average $T_e$ is $\sim 17\mK$ (see Appendix \ref{Annexe:Te}), with minor fluctuations ($\sim0.7$ mK) between experiments and the junction resistance was measured to be 52.5 \Ohm.
Fluctuations in the noise temperature of the amplification chain are canceled out by alternating experimental conditions with references ($I_{\text{dc}}=0$,$I_{\text{ac}}=0$).
Finally, the attenuation along the ac port at $12~\GHz$ ($\sim -79 \dB$) is estimated by fitting the photo-excited auto-correlation, \cite{simoneau_mesures_2021}.

\paragraph{Nonlinearities.}
The statistics of the voltage fluctuations generated by the sample are separated from that coming from the measurement setup by exploiting the linearity of cumulants for statistically independent variables, $\cumulant{\left(X+Y\right)^k} = \cumulant{X^k} + \cumulant{Y^k} $
\cite{virally_discrete_2016,simoneau_photon_2017,simoneau_mesures_2021,simoneau_photocount_2022}.
This holds true only if the amplification chain is linear enough, as nonlinearities introduce artificial correlations between the noise of the sample and that of its detection. This effect worsens the higher the order of the cumulant, and it is much preferable to minimize nonlinearities before acquisition than to rely on error-prone post-processing corrections.
In our experiment, the main source of nonlinearity was traced down to the A/D converter of the digitizer (see Appendix \ref{Annexe:Nonlin}) and was mitigated by reducing by $6\dB$ the gain in the digitizer itself, i.e. we reduce the width of voltage distribution seen by the digitizer by half, essentially using 9 of the 10 available bits.
This sacrifice in dynamic range for linearity effectively resolved all nonlinearity-related issues in measurements, rendering further corrections unnecessary for the presented data.

\subsection{ Numerical methods }
The $32~\GSaPerSec$ voltage samples are numerically converted to photon quadrature using a homemade C++  implementation of the overlap-add method using FFTW for high performance hardware specific FFTs and parallelization on a 36 cores (Dual Xeon E5-2697v4) server \cite{code_Simon}. Results shown herein require long averaging (for example,  Fig. \ref{fig:Fano} took $\sim2$ weeks of continuous averaging), because i) the sample noise represents only between  0.25 and 5\
Maximizing throughput helps both for faster experimental iterations and easier compensation for fluctuations in the setup properties (noise temperature fluctuation of the cryogenic amplifier, electrical noise from the power lines, etc.). The numerical convolutions with kernels of 257 points (number of time bins in $k\conv\beta(t)$) runs at about $2.0~\GSaPerSec$ (1 sample is 10 bit) and a typical experimental loop (setting parameters, acquisition, transfer, numerical treatment) runs at an effective sample rate of  $1.5~\GSaPerSec$.

The digitized signal is deconvolved to remove the effect of the transfer function from the sample to the digitizer,
\begin{eqnarray}
	x(t) 
	&=&
	\underbrace
	{
		k \conv \beta \conv |g|^{-1}
	}_\text{numerical}
	\conv
	\underbrace
	{ 
		g \conv [ v_s(t) + v_A(t) ]  
	}_\text{physical}
	~~~,~
	\label{eq:time_quad_numerical}
\end{eqnarray}
where $k$ identifies the kernels transforming $\v \to \x$, $g$ is the voltage gain between the sample and the digitizer, $v_s$ is the sample's voltage and $v_A$ is the voltage fluctuation added by the detection.
No phase information is contained in our measurement of the power gain, hence we only deconvolve for the modulus of the transfer function, $|g(f)|$. 
This is enough as long as we consider only even order cumulants that only involve modulus squared of the voltage fluctuations, see below.

A numerical kernel $k \conv \beta \conv |g|^{-1}$ is constructed for each mode during calibration and is reused during the experiment for each experimental condition. 
It is noteworthy that since $k\sim 1/\sqrt{|t|}$ contains poles both at $ k(t=0)$ and $k(f=0)$ it needs to be regularized, see appendix \ref{Annexe:k_regular}. As pointed out in \cite{Virally2018}, $k$ is not causal: $x(t)$ depends of future times. This is the price to pay to extract photon statistics from traces of the time-dependent voltage.

\section{Results}
\subsection{Bichromatic thermal radiation}
We first test our experiment by measuring the photocount statistics of thermal radiation in a bichromatic single photonic mode, at 4GHz and 8GHz. 
The radiation comes from the current noise of the tunnel junction when dc biased. 
While this noise is not Gaussian, cumulants of order 3 and above are very small so the radiation generated by the junction is almost identical to the thermal one. 
Changing the dc bias on the junction increases the average photon number, like increasing the temperature does, but the link between the variance and average photon numbers remains the same: $\expval{\dn^2} = \expval{\n}(\expval{\n}+1)$. 

\begin{figure}[!ht]
	\centering
	\includegraphics[width=\linewidth]{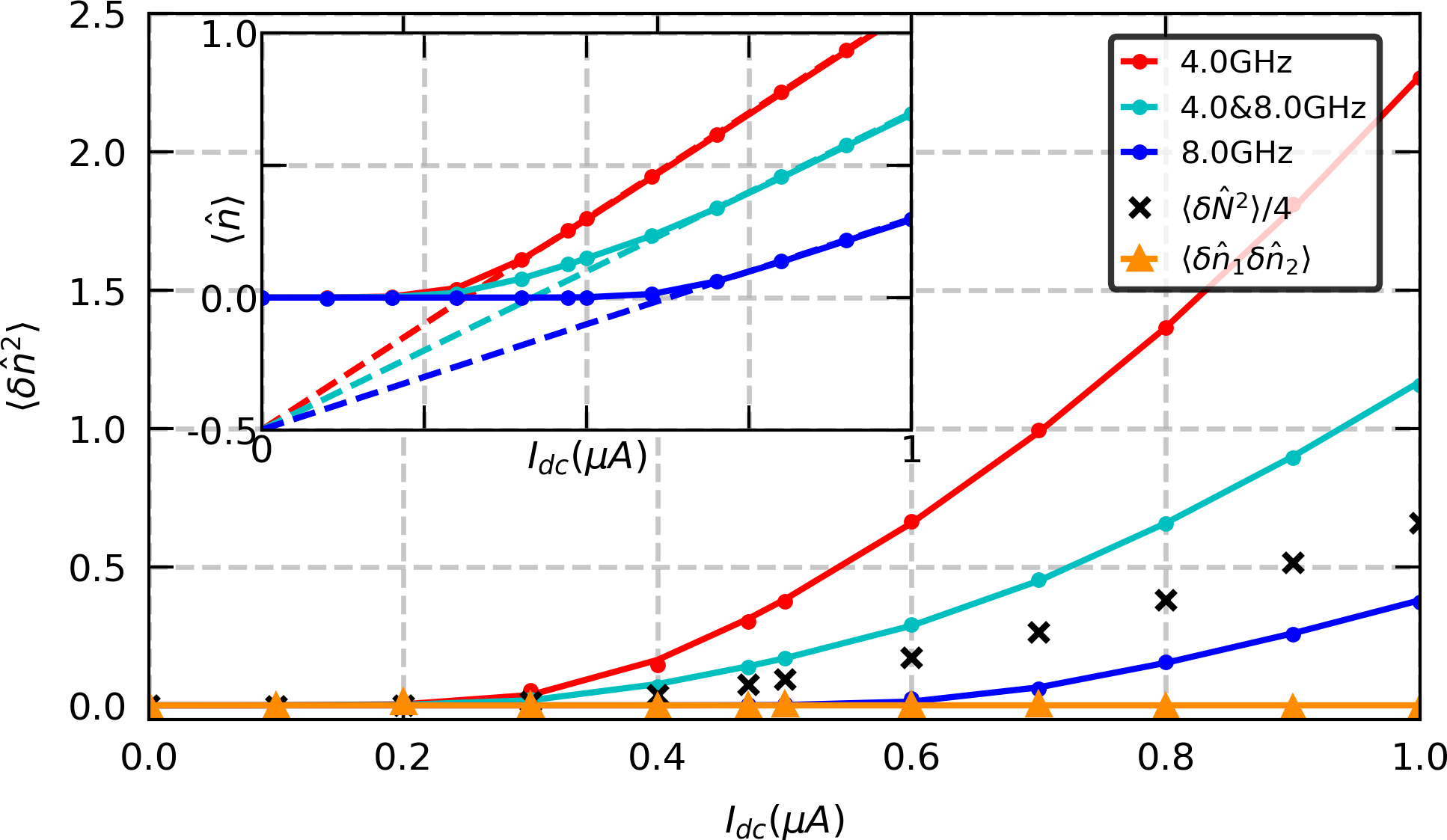}
	\caption{
        Variance of the photon number vs. dc current in the junction. Red (resp. blue) '.' markers correspond to monochrome modes at frequency $f_1=4~\GHz$ (resp. $f_2=8~\GHz$); cyan '.' markers correspond to the bichromatic mode at $4\&8~\GHz$.
        black 'x' markers correspond to $\langle\delta\N^2\rangle/4$ with $\N$ the total number of photons at frequencies $f_1$ and $f_2$; orange '$\triangle$' markers correspond to the correlation $\expval{\delta\n_1\delta\n_2}$.
        Solid lines are theoretical predictions : $\expval{\n}(\expval{\n}+1)$. Inset: average photon number vs. dc current in the junction.
        By extrapolating the linear shot noise regime to the origin (dashed lines) we get the value in photons for vacuum fluctuations ($1/2$ photon for any mode), $\n + 1/2 = \x^2 + \p^2$.  
	}
	\label{fig:moments}
\end{figure}

This is true provided the statistics are performed on a single mode, irrespective of this mode being narrow-band or not. 
In Fig. \ref{fig:moments} we show the measurements of $\expval{\n}$ (inset) and $\expval{\dn^2}$ as a function of the dc bias current in the junction for the monochromatic modes $\a_1$ at frequency $f_1=4~\GHz$ (red markers), $\a_2$ at $f_2=8~\GHz$ (blue) as well as for the bichromatic mode $\b=(\a_1+\a_2)/\sqrt2$ that encompasses both frequencies (cyan). 
The average photon number in the mode $\beta$ emitted by the junction is given by \cite{th_squeezing_inout}:
\begin{equation}
    \expval{\n}=2\int_0^{+\infty} |\beta(f)|^2\left(\frac{S_2(f)}{R hf}-\frac{1}{2}\right) \dd{f}
    \;\; ,
\end{equation}
with $S_2(f)$ the current noise spectral density of the junction at frequency $f$. The factor two in front of the integral is inserted since $S_2(f)$ is usually defined for positive and negative frequencies.  
Solid lines for $\n_1$ and $\n_2$ in the inset of Fig.\ref{fig:moments} correspond to this formula. 
The average number of bichromatic photons is obviously given by $\expval{\n}=(\expval{\n_1}+\expval{\n_2})/2$ (cyan solid line). 
Solid lines in Fig. \ref{fig:moments} correspond to $\expval{\n}(\expval{\n}+1)$. 
They closely follow our experimental data even in the bichromatic case, indicating that we are correctly counting bichromatic photons. 

Let us now consider the operator $\N/2=(\n_1+\n_2)/2$, the mean of the photon number in modes 1 and 2. 
Its average value is the same as that of the single mode photon number $\hat n$. 
Its variance is given by:
\begin{equation}
\frac{1}{4} \expval{\delta\hat{N}^2}=\frac14\left(\expval{\delta\hat{n}_1^2}+\expval{\delta\hat{n}_2^2} \right)
+\frac12 \expval{\delta\hat{n}_1\delta\hat{n}_2}
\;\; .
\label{eq:dN2}
\end{equation}
The last term, $\expval{\dn_1\dn_2}$, corresponds to correlations between photons emitted at different frequencies. 
There are no such correlations in thermal light, as confirmed experimentally, see the orange '$\triangle$' markers in Fig.\ref{fig:moments}.
Then Eq.(\ref{eq:dN2}) reduces to the fact that if two random variables are uncorrelated, their variances add. 
The measurement of $\expval{\delta\hat{N}^2}/4$ is plotted as  black 'x' markers in Fig. \ref{fig:moments}: it clearly differs from $\expval{\dn^2}$ and does not obey the photon statistics of thermal light. 
The difference between the two comes from the operator $\hat\Delta$ introduced in Eq.(\ref{eq:Delta}). $\hat\Delta(t)$ oscillates at frequency $f_2-f_1$ and is zero in average. Thus it does not contribute to the average photon number. 
However $\expval{\hat\Delta^2}\neq0$ contributes to $\expval\dn^2$ to make it obey the statistics of thermal light. 
This term reflects the fact that electromagnetic field adds, $\b\propto \a_1+\a_2$, not the photon numbers, $\n\neq\n_1+\n_2$.

\subsection{  Bichromatic single mode squeezed radiation }
In the presence of an ac excitation, the tunnel junction emits a two-mode squeezed radiation by noise modulation  \cite{gasse_observation_2013,bbrb2013}. 
We consider the case where a dc voltage $\Vdc =hf_p/(2e)$ is superimposed on a sine wave at frequency $f_p=12~\GHz$. 
In these conditions the junction emits pairs of photons at frequencies $f_1$ and $f_2$ such that $f_1+f_2=f_p$ (three-wave mixing) where $f_1$ and $f_2$ can take any value between 0 and $f_p$\cite{forgues_non-classical_2016}. 
Hereunder we consider the counting statistics of photons in the bichromatic mode $b$, as above. 
If the junction emitted pure squeezed vacuum, one would have the photocount variance given by $\expval{\dn^2}=2\expval{\n}(\expval{\n}+1)$, i.e. the double of that of thermal light\cite{barnett_methods_2002}. 
The junction, however, does not emit only pairs of photons since $\expval{\n_1}\neq\expval{\n_2}$. 
The photocount variance has been measured in the limit $f_1\simeq f_2$ where the two sidebands were not resolved, i.e. in the limit of single-mode squeezing\cite{simoneau_photon_2017}. 
Here we extend these results to well separated frequencies $f_1=4$GHz and $f_2=8~\GHz$. 
We characterize $\expval{\dn^2}$ by defining the Fano factor $\mathcal{F}=\expval{\dn^2}/\expval{\n}$. For thermal radiation, $\mathcal{F}=\expval{\n}+1$.

\begin{figure}[!ht]
	\includegraphics[width=\linewidth]{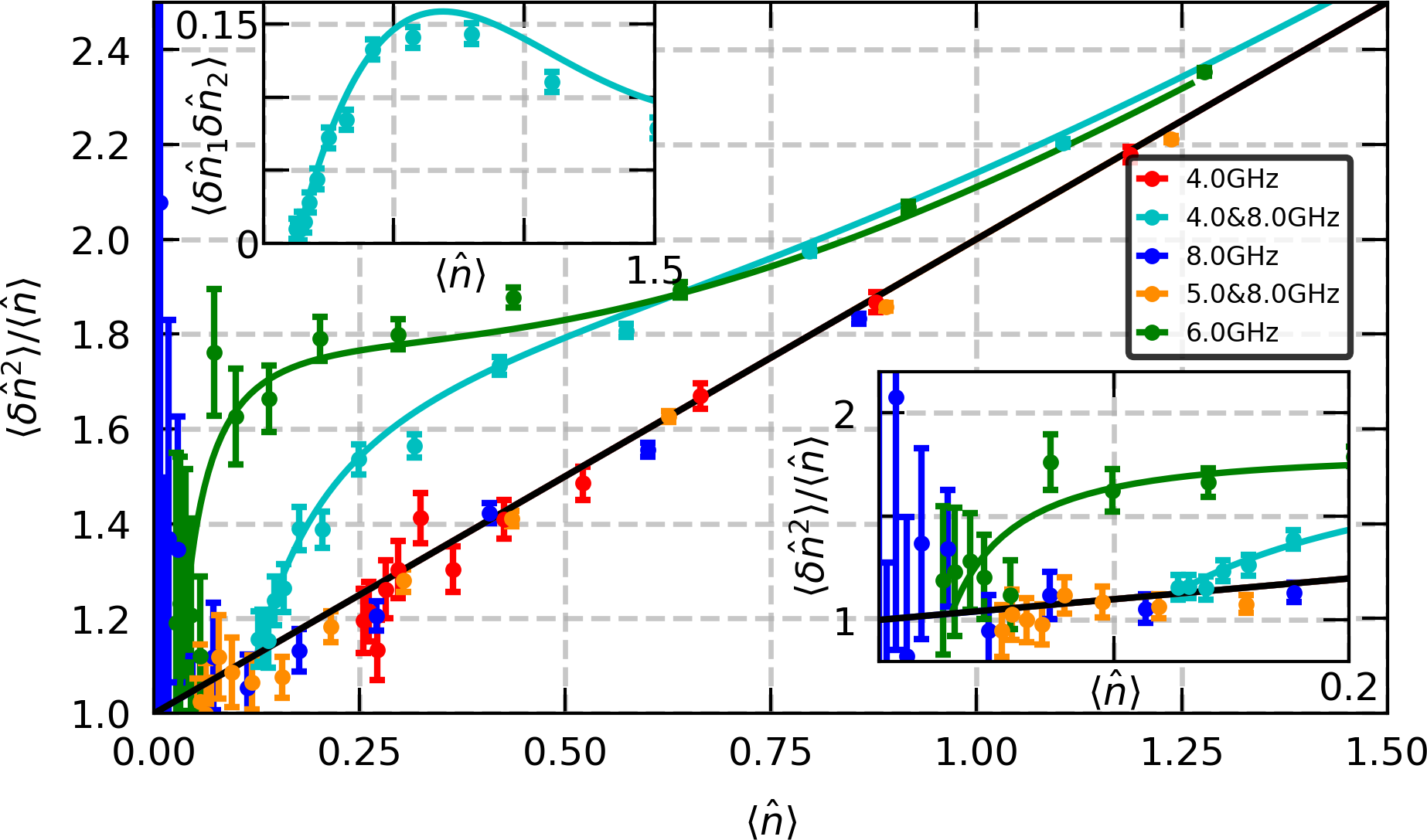}
	\caption{The Fano factor as a function of the average photon number for different photonic modes. Markers are experimental data, solid lines theoretical predictions. Thermal state (solid black line) i.e. $\mathcal{F}=(\expval{n}+1)$.
    Theory for monochromatic modes at 4 GHz (red), 8 GHz (blue) as well as for the 5\&8 GHz mode (orange) corresponds to thermal light. The average photon number is varied by varying the amplitude of the ac drive at frequency $f_p=12$ GHz with a constant dc bias $\Vdc=hf_p/(2e)$. 
    Upper left inset: Correlation between photon numbers at 4 and 8 GHz vs. average bichromatic photon number.
    Lower right inset: A close-up view of the leftmost part of the graph. 
    }\label{fig:Fano}
\end{figure}

We show in Fig. \ref{fig:Fano} Fano factors vs. $\expval{\n}$ for various amplitudes of the ac drive and various detection modes. 
If one detects only one frequency, such as 4 GHz (red) or 8 GHz (blue) which do not contain pairs of photons, one obtains thermal statistics (black), as well known for two-mode squeezed light\cite{barnett_methods_2002}. 
A bichromatic mode made of two frequencies that do not add up to 12\GHz, such as 5 GHz and 8 GHz follow the same statistics (orange). 
The result for a monochromatic detection at 6 GHz (green) shows a clear deviation from $\expval{\n}+1$. 
It corresponds to the result reported in \cite{simoneau_photon_2017} with a lower electronic temperature. 
The cyan markers correspond to the case $f_1=4~\GHz$, $f_2=8~\GHz$: this bichromatic mode indeed bears the signature of the correlations existing between the electromagnetic fields at frequencies $f_1$ and $f_2$, characteristic of two-mode squeezed radiation. 
Theoretical expectation for squeezed condition, $6~\GHz$ (green) and $4\&8~\GHz$ (cyan), are computed from the variance and fourth cumulant of current fluctuations as described in \cite{forgues_non-classical_2016,th_squeezing_inout}.

The photocount variance can be rewritten as:
\begin{equation}
    \expval{\dn^2}=\expval{\n}(\expval{\n}+1)+\expval{\dn_1\dn_2}
    \;\; .
    \label{eq:dn2_nnp1}
\end{equation}
As clear from this expression, the deviation from the statistics of thermal light uniquely comes from correlations between photons at frequencies $f_1$ and $f_2$. In the case of pure squeezed vacuum, only pairs are emitted so $\expval{\n}=\expval{\n_1}=\expval{\n_2}$ and $\expval{\dn_1\dn_2}=\expval{\dn_1^2}$, i.e. the Fano factor is doubled as compared to thermal light. In our experiment the junction emits pairs of photons \cite{forgues_emission_2014}, see the inset of Fig. \ref{fig:Fano}, but not only.

The two measurements discussed until now were meant to demonstrate the ability of our experiment to measure the photocount statistics of broadband signals, in particular bichromatic ones with far apart frequencies. In the following we will apply it to explore some of the potential it offers to analyze more generally broadband radiation.  

\subsection{Squeezing spectrum}
Analyzing two-mode squeezed radiation at far apart frequencies may be performed by detecting the variance of two quadratures (in-phase and out-of-phase) at each frequency $f_1$ and $f_2$, i.e. $\X_{1,2}$ and $\P_{1,2}$, which are combined into $\X=(\X_1+\X_2)/\sqrt2$ and $\P=(\P_1+\P_2)/\sqrt2$. 
Correlations between quadratures at different frequencies, for example $\langle\X_1\X_2\rangle$ leads to the joint quadrature $\X$ having a variance below that of vacuum, i.e. $\langle\X^2\rangle<1/2$. 
This procedure usually requires having phase coherence between the detection at $f_1$ and $f_2$ as well as the excitation at $f_p$ that leads to the generation of squeezed vacuum by the sample, whatever it is \cite{Forgues_Bell}. 
In contrast, our technique does not require such stringent experimental constraints and is extremely frequency agile in the sense that analyzing another pair of frequencies only requires changing a single line of code. 
In fact, all the frequencies can be analyzed in parallel with a single acquisition. 
Our method, however, provides photocount statistics, not the variance of the quadratures. 
Below we show how these can be deduced from our measurements, which allows us to measure the squeezing spectrum of a broadband radiation \cite{Wustmann_2013,Grimsmo_2017}.

The quadratures at single frequencies $f_1,f_2$ correspond to the operators $\X_{1,2}=(\a_{1,2}+\adag_{1,2})/\sqrt2$ and $\P_{1,2}=-i(\a_{1,2}-\adag_{1,2})/\sqrt2$. Considering again the bichromatic mode $\b=(\a_1+\a_2)/\sqrt2$, the joint quadratures $\X$ and $\P$ can be expressed as a function of $\b$ and $\bdag$ as : 
$\X=(\X_1+\X_2)/\sqrt2=(\b+\bdag)/\sqrt2$ and $\P=(\P_1+\P_2)/\sqrt2=-i(\b-\bdag)/\sqrt2$. This simply means that the two-mode squeezed state involving the modes $\a_1$ and $\a_2$ is in fact a single-mode squeezed state in terms of $\b$. 

We now show how to deduce the variance of the two quadratures knowing the photon statistics for a mode $\b$ of any frequency content, not only bichromatic.
Introducing the usual operator $\m=(\b^2+\bdag[2])/2$, one has\cite{Grimsmo_2017}:
\begin{eqnarray}
    \n &=& \frac12(\X^2+\P^2-1) \\
    \m &=& \frac12(\X^2-\P^2) \;\; .
\end{eqnarray}
$\m$ being sensitive to a global phase in $\b$, we choose it so $\expval{\b^2}$ is real and positive, i.e. so $\expval{\m}=\expval{\b^2}$.
Supposing that $\b$ is Gaussian and applying Wick's theorem to calculate $\expval{(\bdag\b)^2}$, we find:
\begin{equation}
    \expval{\delta\n^2}=\expval{\n}(\expval{\n}+1)+\expval{\m}^2
\end{equation}
which allows us to deduce $\expval{\m}$ from the photon statistics. $\expval{\m}$ measures the difference between the variance of $\n$ and its value for thermal radiation. 
In the case of the bichromatic mode, and using Eq.(\ref{eq:dn2_nnp1}), one has: $\expval{\m}^2=\expval{\dn_1\dn_2}$, i.e. $\expval{\m}$ is also a measurement of the correlations between photons at frequencies $f_1$ and $f_2$.

\begin{figure}[!h]
    \centering
    \includegraphics[width= \linewidth]{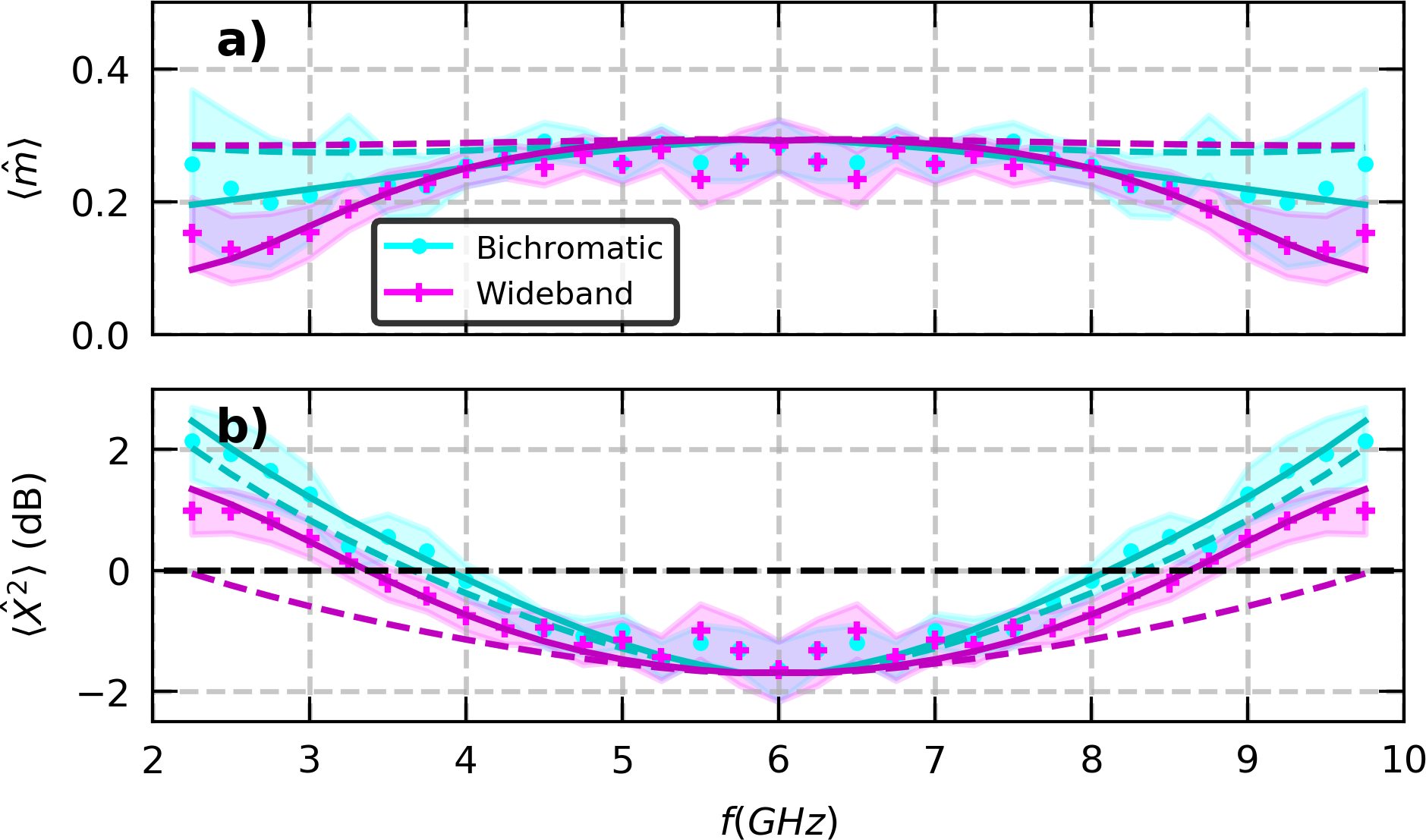}
    \caption{$\expval{\m}$ in linear scale (a) and variance of the quadrature $X$ in dB referred to vacuum (b) as a function of frequency $f$ for a tunnel junction excited at frequency $f_p=12$ GHz with an RMS ac bias of $0.43~\mu$A and $\Vdc=hf_p/(2e)$. 
    Cyan corresponds to bichromatic modes at frequencies $f$ and $f_p-f$ with a width of 200 \MHz. 
    Pink corresponds to wideband photonic modes that extend from $f$ to $f_p-f$. Experimental uncertainty $(3 \sigma)$ is represented as a shaded area around the average curve, '.' or '+' markers. 
    The theory is presented using the same color with dashed lines for a dispersion-less model and solid lines for a theory with a phase that increases quadratically with frequency.
    The black dashed horizontal line at 0dB (bottom graph) represents the vacuum level of fluctuations.
    }
    \label{fig:squeezing_spectrum}
\end{figure}

We now apply our technique to the broadband squeezed radiation generated by the tunnel junction. 
We show in Fig. \ref{fig:squeezing_spectrum} in cyan $\expval{\hat m}$ and the variance $\expval{\X^2}$ of bichromatic modes centered at frequencies $f$ and $f_p-f$ for $f$ varying between 2.25 and 9.75 GHz. 
It exhibits broadband two-mode squeezing  from 4 GHz to 8 GHz. 
The squeezing we obtain is close to that of \cite{esposito_TWPA2022} with a much wider bandwidth, which is similar to that of the most advanced traveling wave parametric amplifiers (TWPAs)\cite{Hakonen_2022,Qiu_2023}. 

The case of "white" photonic modes, which correspond to $\beta(f)$ uniform between $f$ and $f_p-f$, is displayed in pink in Fig. \ref{fig:squeezing_spectrum}. 
These modes exhibit squeezing extending from 3.5 to 8.5 \GHz, starting at $\langle\X^2\rangle\sim -1.6 \pm 0.5~\dB$ for a narrow band detection around 6GHz, reaching $-1.1 \pm 0.3~\dB$ for the wideband mode 5-7 GHz. $\langle\X^2\rangle<0~\dB$ is observed for modes as wide as 3.5-8.5 GHz.
The squeezing obtained with the wideband mode does not go lower than that obtained with bichromatic modes, but it stays at the minimum level on a wider frequency range. This probably stems from these modes containing pairs close to 6 \GHz, which are the most correlated.

Wideband modes have another remarkable property: since they involve a bandwidth much wider than the bichromatic modes, the signal to noise ratio of their measurement is much higher. 
For example, the bichromatic mode at 4 and 8 GHz has a bandwidth (the integrated area under $|\beta(f)|^2$) of 400 MHz while that of the wideband mode 4-8 GHz has a 10 times larger bandwidth, i.e. its measurement requires 10 times less time.

We show in Fig. \ref{fig:squeezing_spectrum} the theoretical expectations as dashed lines. The prediction for $\expval{\m}$ is remarkable, it is almost frequency independent over the whole frequency range: the junction is truly a broadband pair emitter. 
While theory and experiment match between 4 and 8 GHz, there is a discrepancy between both for too far apart frequencies, in particular for wideband modes. We think this is due to frequency dispersion in the detection. 
Phase rotation within the bandwidth of $\beta$ affects the measurement.
For bichromatic modes, a significant phase rotation within 200 MHz matters, but not between $f$ and $f_p-f$. Wideband modes are more sensitive to dispersion. 
Adding an artificial phase rotation that grows quadratically with frequency (a linear increase corresponds to a delay, which has no influence) leads to the solid lines in Fig. \ref{fig:squeezing_spectrum}, which fits our data for a total rotation of $5\pi$ within 1-10 GHz. Dashed and solid cyan lines are indistinguishable, in agreement with the fact that bichromatic modes very weakly sensitive to phase rotation. 
More experiments are needed, with a well-calibrated phase response of the setup, to understand in depth the importance of dispersion in our measurement. 

\subsection{Symmetric  bipartite correlations: entanglement}

\begin{figure}[!h]
    \centering
    \includegraphics[width= \linewidth]{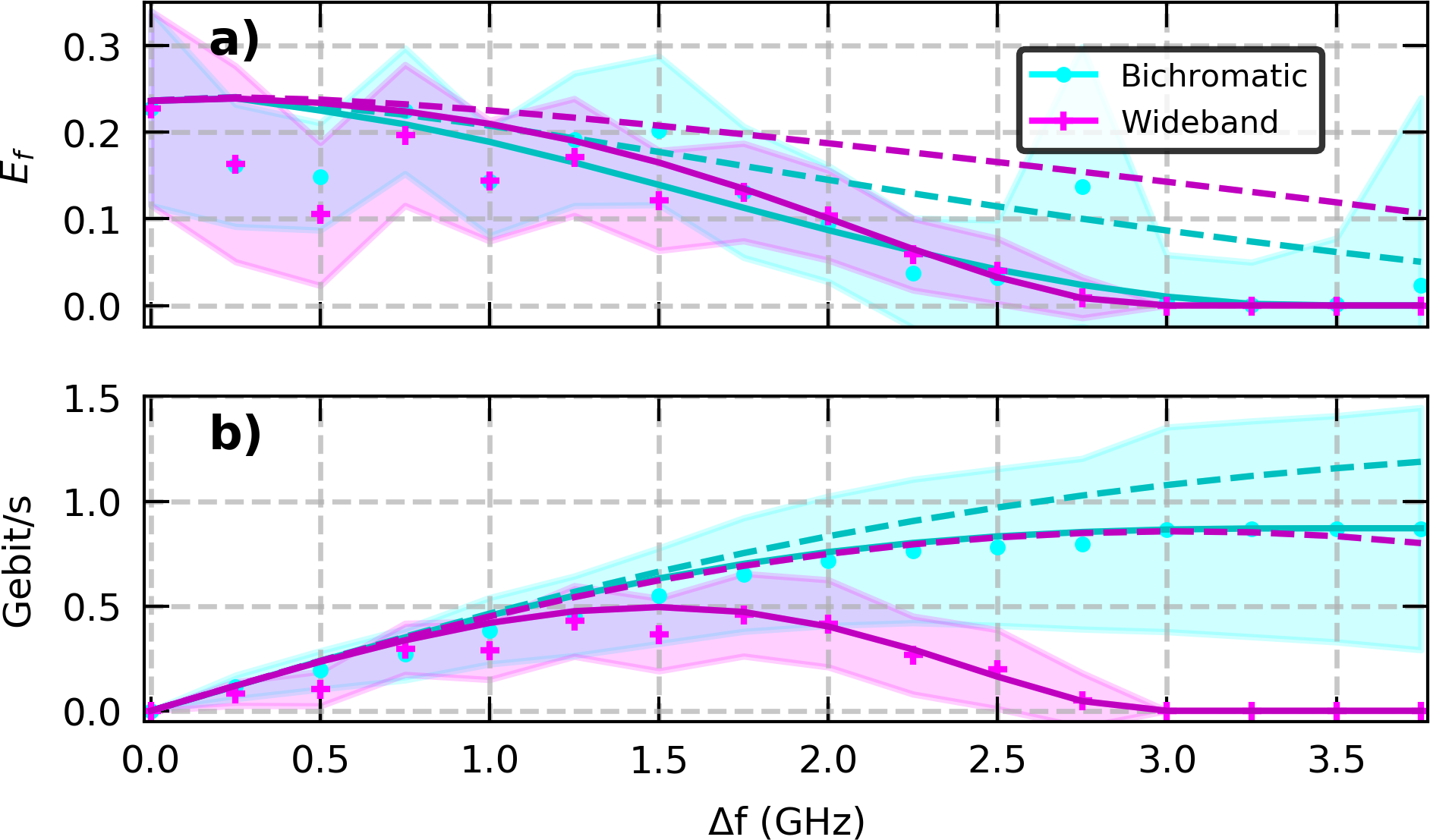}
    \caption{The entanglement of formation $E_f$ in bit per mode (a) and the rate of entanglement (b) in entangled bits per second as a function of difference of frequency $\Delta f$ from 6 GHz.
    Conditions are the same as in Fig. \ref{fig:squeezing_spectrum}.
    The theory is presented using the same color with dashed lines for a dispersion-less model and solid lines for a theory with a phase that increases quadratically with frequency.
    Experimental uncertainty $(3 \sigma)$ is represented as a shaded area around the average curve, '.' or '+' markers. 
    }
    \label{fig:Ef_Erate}
\end{figure}

As we show below a tunnel junction is an interesting candidate for the generation of entanglement or steering, and our measurement method an efficient way to measure such relevant quantities.
We first compute the entanglement formation $E_f$ and entanglement rate, in e-bits/s. 
For this we consider a fictitious bipartite configuration where detector $A$ has access to frequencies below $F=6$ GHz and detector $B$ frequencies above $F$. 
Note that $A$ and $B$ could be spatially separated by the use of a diplexer, a dispersive device that does not add any noise. Bichromatic modes correspond to the usual configuration where $A$ and $B$ receive quasi-monochromatic signals. 
In contrast, for wideband modes $A$ and $B$ each receive broadband (up to $\sim3$ GHz) signals. $E_f$ is computed as usual from the symplectic covariance matrix between $A$ and $B$, simply related to the statistics of $\expval{\hat n_A}$, $\expval{\hat n_B}$ and $\expval{\hat m}$ \cite{Adesso_2005,Adesso_2014, He_2015}. 
The result is presented in Fig. \ref{fig:Ef_Erate} (a) as a function of $\Delta f$. 
For bichromatic modes $\Delta f$ is half the difference between the frequencies of $A$ and $B$ while for wideband modes it represents their bandwidth. We observe an entropy of formation of $0.23$ at low $\Delta f$. As a comparison, a pure 2dB squeezed vacuum corresponds to $E_f=$ 0.3 . 
While $E_f$ decreases for increasing $\Delta f$, it appears that the theoretical $E_f$ decays much slower for wideband modes than for bichromatic ones. 
Note, however, that benefiting from the entanglement rate of bichromatic modes requires working with  many detection modes, one per pair of frequencies.

The entanglement rate for bichromatic modes is calculated by integrating $E_f$ over frequency \cite{esposito_TWPA2022, Hakonen_2022}, and is shown in Fig. \ref{fig:Ef_Erate} (b). 
It starts by increasing linearly with the bandwidth, then saturates since increasing the bandwidth does not add entangled bits. 
Wideband modes are single photonic modes with a bandwidth $2\Delta f$. 
They carry an entanglement rate of $2E_f\Delta f$. This grows linearly with $\Delta f$ and is equal to the entanglement rate of bichromatic modes at low $\Delta f$, but goes to zero at too large $\Delta f$ since $E_f$ decays faster than $1/\Delta f$ at large bandwidth.     

In Fig. \ref{fig:Ef_Erate} we report an entanglement rate of 0.87 Gebit/s (1.19 theoretical maximum) for bichromatic modes, and 0.46 Gebit/s (0.86 theoretical maximum) for wideband modes. 
By achieving stronger squeezing, the best reported entanglement rate in TWPAs is 2 Gebit/s \cite{Hakonen_2022}.

While a tunnel junction provides less squeezing than the best TWPAs, it should be emphasized that its fabrication is by far less involved than that of a TWPA. 
Increasing the squeezing in tunnel junctions could be achieved by shaping the time-dependent excitation, with a predicted bound of -3.9 dB obtained for a periodic excitation of Dirac voltage pulses \cite{Mendes_2015}.
However we show in Appendix  \ref{Annexe:Eof_sqvac} that $E_f$ for a two-mode vacuum squeezed state grows at best logarithmically with the squeezing intensity. 
Another option to increase the entanglement rate is to increase the bandwidth of the signal. 
The bandwidth of a tunnel junction is only limited by its RC time, here $>15$ GHz.  
It can reach 100 GHz by working with a more transparent $\mathrm{AlO_x}$ barrier and above 200 GHz using AlN barrier\cite{Lodewijkb2009}. 

\subsection{Asymmetric  bipartite correlations: steering}

In the separation of the noise generated by the junction into two detectors $A$ and $B$ introduced above, both detectors are not experiencing similar signals: there are more photons emitted at low frequency than at high frequency, i.e. $\expval{\hat n_A}>\expval{\hat n_B}$. This natural asymmetry exactly corresponds to the situation where quantum steering could be looked for.
\begin{figure}[!h]
\centering
\includegraphics[width= \linewidth]{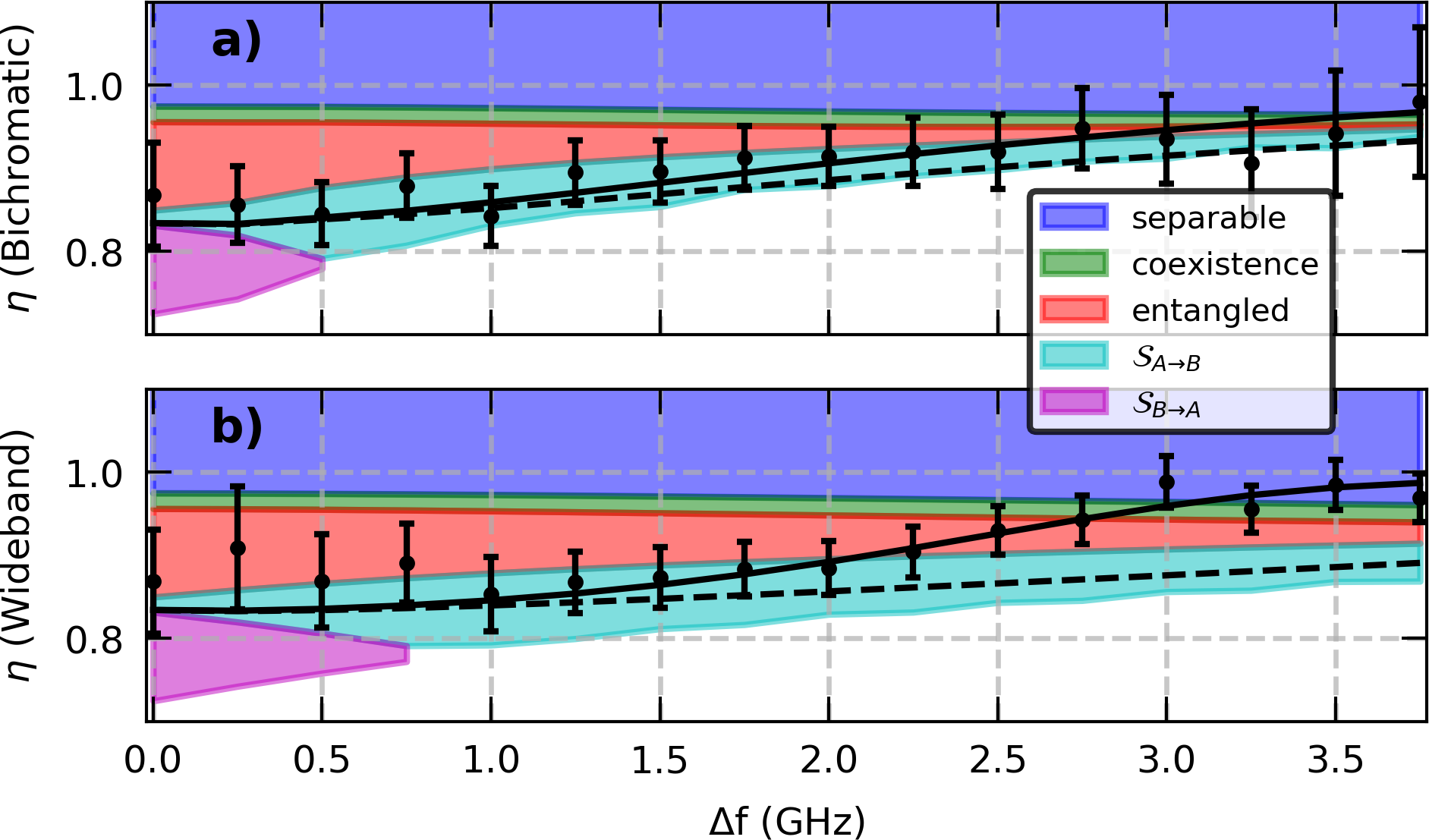}
\caption{ The measured parameter $\eta$ (black markers) as a function of the difference of frequencies $\Delta f$ from 6 GHz for bichromatic states (a) and wideband states (b). Colored areas correspond to the criteria for separability and steering of bipartite Gaussian states according to \cite{Kogias_2015}. The tunnel junction is excited at frequency $f_p=12$ GHz with an RMS ac bias of $0.34~\mu$A and $\Vdc=hf_p/(2e)$, different from the bias that optimizes squeezing.}
\label{fig:Steering_11}
\end{figure}

In \cite{Kogias_2015} are given a series of criteria to know whether a bipartite Gaussian state is separable, partially entangled or steerable, from $A$ to $B$ ($\mathcal{S}_{A\to B}$) or from $B$ to $A$ ($\mathcal{S}_{B\to A}$). 
These criteria require the evaluation on a single parameter $\eta=(\mu_A\mu_B)/\mu$ with $\mu$ the purity of the global state, and $\mu_A$, $\mu_B$ those of the states detected by $A$ and $B$ respectively. 
Our detection technique allows us to calculate the purities, and so the factor $\eta$ as well. 
We report in Fig. \ref{fig:Steering_11} $\eta$ vs. $\Delta f$ for bichromatic and wideband modes. 
Various colored regions correspond to the above-mentioned criteria. 
We clearly observe steering in the noise of the junction for bichromatic modes up to $\sim6$ GHz apart and wideband modes up to $4$GHz wide.
In the absence of phase dispersion, steering would theoretically be possible at all $\Delta f$ considered (dashed black lines in Fig. \ref{fig:Steering_11}(b)). 
A previous attempt to see steering in the noise of a tunnel junction had led to a negative result \cite{Forgues_Bell}. 
Despite a similar electronic temperature, here we clearly observe steering.
This is for 3 reasons. First, wider modes (higher $\Delta f$) steer more easily as the theory shows. 
Second, it seems easier to violate the steering criterion at a lower ac current than the optimal current for squeezing. The reason for this is not trivial since the criteria and $\eta$ are both functions of the purities, which in turn depend on the ac voltage \cite{Kogias_2015}.
In Fig. \ref{fig:Steering_11} the junction was ac biased by an RMS current of $0.34~\mu$A, while in Fig. \ref{fig:squeezing_spectrum}, $I=0.43~\mu$A.
For the sake on completeness Fig. \ref{fig:Steering_13} in appendix \ref{Annexe:Steering_13} shows the measured steering in the same biasing conditions than in Fig. \ref{fig:squeezing_spectrum}: less steering is observed in these conditions. 
Finally, we used the more general criterion developed in \cite{Kogias_2015} valid for arbitrary bipartite Gaussian states.
Steering is also observed for wideband states which will present higher key rates \cite{Kogias_2015} for the same level of monomode bichromatic steering, similarly to what is observed regarding $E_f$ and entanglement rate.

Steering (measured by a relation on purities) being a type of asymmetric quantum correlation and entanglement (measured by a going below vacuum) being symmetrical, it is natural to ask the question whether a quantitative measure of both of them would be maximal in the same experimental conditions.
This is beyond the scope of this article but should be explored in future work.

\subsection{Asymmetric modes and criterion for inseparability}

Until now we have considered bichromatic modes with
equal weights on the two frequencies $f_1$ and $f_2$. We now focus on
the effect of the relative weights of two modes $a_1$ and $a_2$by  defining the combined mode:
\begin{equation}
    \hat d(\lambda)=\sqrt{1-\lambda}\,\a_1+\sqrt\lambda\,\a_2
    \;\;\; ,
    \label{eq:lambda}
\end{equation}
which goes from $\hat d(\lambda=0)=\a_1$ to $\hat d(\lambda=1)=\a_2$. Adding a complex phase on either of the two terms does not influence the photon statistics. As above, we consider the effect of $\lambda$ on the two quantities, $\expval{\m}$ and $\langle\X^2\rangle$.
These considerations are reminiscent to what has been introduced by Duan et al. to construct a criterion for inseparability \cite{Duan_2000}. In this work, two EPR-like operators $\hat u$ and $\hat v$ are constructed from the quadratures $\hat x_{1,2}$ and $\hat p_{1,2}$:
\begin{eqnarray}
        \hat{u} &=& |a| \x_1 + (1/a) \x_2 \\
        \hat{v} &=& |a| \p_1 - (1/a) \p_2
\end{eqnarray}
with $a$ a real parameter. Taking $\lambda =a^4/(1+a^4)$, the broadband quadrature $\hat X$ related to $\hat d$ reads $\hat X=(\sqrt\lambda/a)\hat u$. The definition of $\hat v$ contains a minus sign while that of $\hat P$ does not. For a covariance matrix that is symmetric, i.e. $\expval{\x_1\x_2}=-\expval{\p_1\p_2}$,  one has $\expval{\hat u^2}=\expval{\hat v^2}$ so the criterion of Duan et al. (Eq.(3) in \cite{Duan_2000}) is simply:
\begin{equation}
    \expval{X^2} \geq \frac{1}{2}.
\end{equation}
Thus the modes $\a_1$ and $\a_2$ being inseparable is equivalent to the combined mode $\hat d$ being squeezed for certain values of $\lambda$. This result is not specific to narrow-band modes, it is valid for any two modes $a_1$ and $a_2$. For example, $a_1$ could cover frequencies below $F$ and $a_2$ above $F$, as introduced for the bipartite correlations discussed above.

\begin{figure}[!ht]
	\centering
	\includegraphics[width=\linewidth]{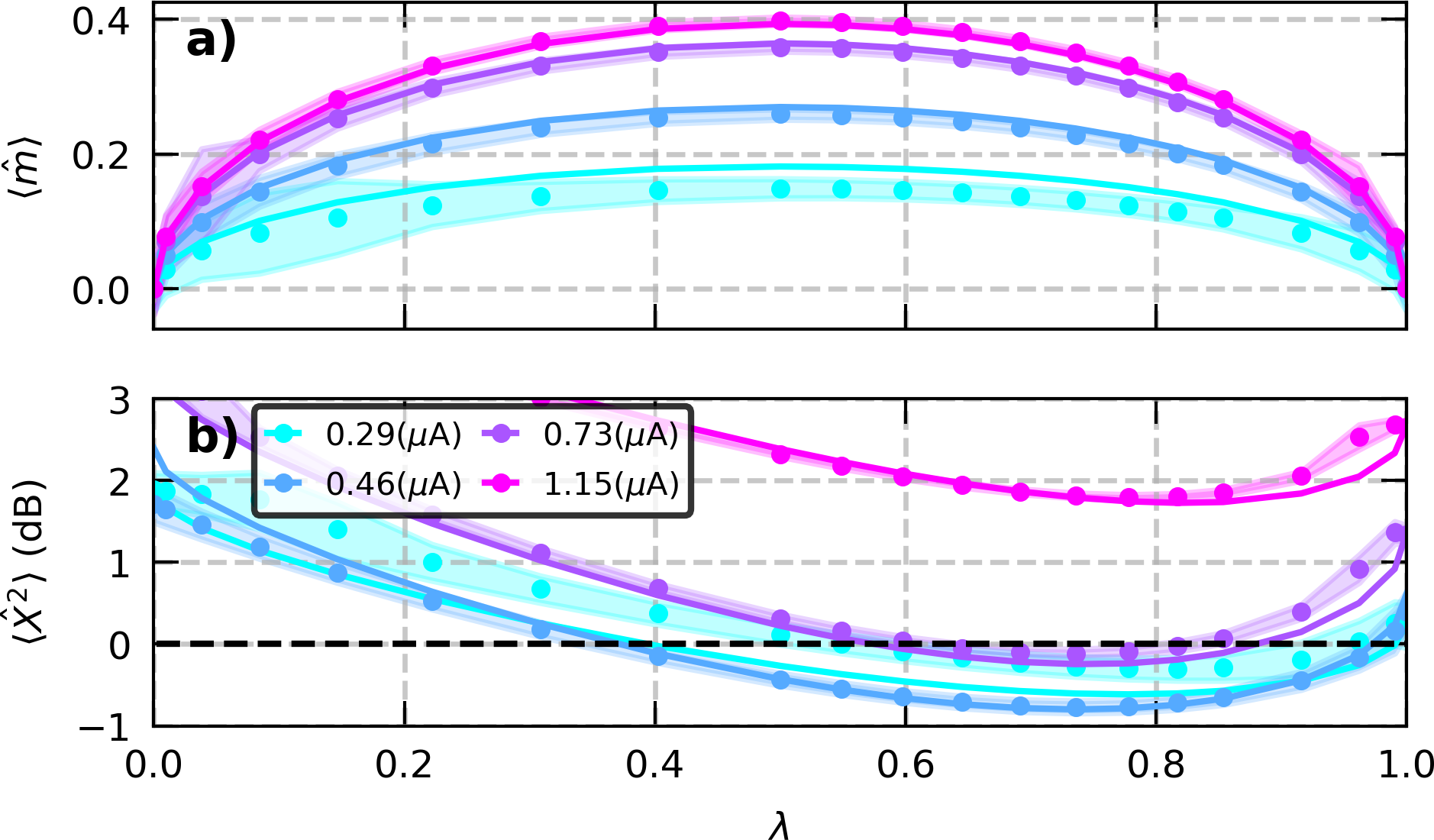}
	\caption{$\expval{\m}$ in linear scale (a) and variance of the quadrature $X$ in dB referred to vacuum (b) as a function of the parameter $\lambda$, see Eq.(\ref{eq:lambda}), with $f_1=4.25~\GHz$ and $f_2=7.75~\GHz$ and various ac excitations. Experimental uncertainty $(1 \sigma)$ is represented as a shaded area around the average curve. 
    Markers are experimental data and solid lines are theory.
    The black dashed horizontal line represents the vacuum level of fluctuations.
    }
	\label{fig:lambda_on_X2}
\end{figure}

Below we show experimentally by varying $\lambda$ that squeezing may be optimal away from $\lambda=0.5$, i.e. for modes $d$ with asymmetric weights. We focus on bichromatic modes with fixed frequencies $f_1=4.25$ GHz and $f_2=7.75$ GHz and show in Fig. \ref{fig:lambda_on_X2}a $\expval{\hat m}$ vs. $\lambda$ for various ac currents in the junction. For $\lambda=0$ and $\lambda=1$, one has $\expval{\hat m}=0$ since there is no pair in the detection mode. 
The maximum is obtained for $\lambda=0.5$. Indeed:
\begin{eqnarray}
    \expval{\hat m}
    &=&
    4\sqrt{\lambda(1-\lambda)}\mathrm{Re}\expval{\a_1\a_2}
    \;\; .
\end{eqnarray}

This is sound since a photon at frequency $f_p$ gives rise to a pair of photons, one at frequency $f_1$ and one at frequency $f_2$. The optimal detection is the one that respects this symmetry.

We show in Fig.\ref{fig:lambda_on_X2}b $\langle\X^2\rangle$ as a function of $\lambda$ for various ac currents in the junction. For $\lambda=0$ or $\lambda=1$ there is no squeezing as no photon pair is detected, thus $\langle\X^2\rangle>1/2$. 
Since $f_2>f_1$, $\expval{\n_1}>\expval{\n_2}$, thus $\langle\X^2\rangle$ is larger for $\lambda=0$ than for $\lambda=1$, as observed. 
In between there might be squeezing, as shown in Fig.\ref{fig:squeezing_spectrum} which corresponds to $\lambda=0.5$. 
This value, however, does not correspond to the minimum, which is clearly closer to $1$ than to $0$. 
This comes from the fact that $\expval{\n_1}\neq\expval{\n_2}$: the junction emits not only pairs, it also emits single photons, and more at frequency $f_1$ than $f_2$. 
Thus the minimum of $\langle\X^2\rangle$ results from a compromise between having as many pairs as possible, i.e. $\lambda=0.5$ and having as little single photons as possible, i.e. $\lambda=1$. 
More precisely: 
\begin{eqnarray}
    \langle\X^2\rangle
    &=&
    \frac12+(1-\lambda)\expval{\n_1} + \lambda\expval{\n_2} 
\nonumber\\
    &-& 4\sqrt{\lambda(1-\lambda)}\mathrm{Re}\expval{\a_1\a_2} \;\;.
\end{eqnarray}

For wideband modes, $\expval{\m}$ and $\langle\X^2\rangle$ depend on the shape of the mode $\beta(f)$, both its amplitude and phase. 
Finding the wideband mode that optimizes $\expval{\m}$ or $\langle\X^2\rangle$ is a captivating but daunting task that goes beyond the scope of this article. 
The same problem has been addressed with fermionic signals for which a procedure has been found \cite{Roussel_2021}. 
Its extension to the case of a bosonic field that we are dealing with here would be of the utmost interest. 

\section{Conclusion}

We have reported an experiment able to access the photon statistics of single modes with a frequency content that may span from 1 to 10 GHz. 
From this we showed how to deduce squeezing spectra and how to adjust the choice of bichromatic modes for optimal detection of squeezing. 
By applying our technique to the broadband radiation emitted by a tunnel junction, we demonstrate squeezing over a bandwidth from 4 to 8 GHz, and show very high entanglement rate (0.87 Gebit/s measured, up to 1.19 Gebit/s achievable).
We also reported the existence of quantum steering in the noise emitted by a tunnel junction.
We demonstrated how the presence of squeezing in any mode is related to the presence of entanglement and how most of this entanglement can be captured by a well-chosen wideband single mode.

Our technique can be applied to a large variety of wideband sources, from superconducting traveling wave parametric amplifiers (TWPAs) \cite{shu_nonlinearity_2021, macklin_nearquantum-limited_2015, Grimsmo_2017, fasolo_josephson_2021, esposito_perspective_2021, esposito_TWPA2022,Hakonen_2022,Qiu_2023} to maybe the radiation generated by quantum Hall conductors \cite{Bartolomei_2023}. 
Other applications are those based on a dc voltage-biased Josephson junctions, for which there is no ac excitation, so no way to measure directly the quadratures \cite{Jebari_2018, Rolland_2019, Peugeot_2021, Menard_2022, Albert_2024}.  

Our experiment opens the possibility to deal not only with pairs of frequencies but with ultra-broadband signals in general. 
While in the parametric downconversion process, it is quite clear that one photon at frequency $f_p$ gives rise to a pair at frequencies symmetric around $f_p$, it is not clear at all that the mechanism of noise modulation involved in electronic noise is so simple. 
For example, does the junction emit randomly pairs of different frequencies, or broadband pulses which contain correlation among frequencies? In the latter case there could be correlations between different pairs. Our setup is the perfect tool to answer such questions, which could provide a new viewpoint on electronic quantum transport.

\begin{acknowledgments}
We are very grateful to Stéphane Virally for the many fruitful discussions, and to Gabriel Laliberté and Christian Lupien for their technical help. This work has been supported by the Canada Research Chair program, the NSERC, the Canada First Research Excellence Fund, the FRQNT, and the Canada Foundation for Innovation.
\end{acknowledgments}

\bibliography{bicolor}

\begin{thebibliography}{52}
\bibitem [{Our custom python librarie ( C++ librarie exported to python ) to compute time domain quadratures is available at : github.com/SimonBolducBeaudoin/Time\_quadratures()}]{code_Simon}%
  \BibitemOpen
  Our custom python librarie ( C++ librarie exported to python ) to compute time domain quadratures is available at : github.com/SimonBolducBeaudoin/Time\_quadratures,\ \href@noop {} {}\BibitemShut {NoStop}%
\end{thebibliography}

\clearpage

\appendix

\section{Numerical convolution ($\v(t)\to\x(t)$)}
\label{Annexe:k_regular}

In order to implement the convolution \eqref{eq:convolution}, we need a numerical representation for the kernel $k(t)$  that properly handles the pole at $t=0$.
This is done by using the fact that the process of digitizing the signal with time steps $\Delta t$ implies low-pass filtering at Nyquist frequency $f_N=1/(2\Delta t)$.

We use:
\begin{eqnarray}
    \frac{1}{\sqrt{|t|}} &=&  \int_0^{+\infty} \frac{2 \cos(2\pi f t)}{\sqrt f} \dd f
    \nonumber\\
    &\approx&  \int_0^{f_N} \frac{2\cos(2\pi f t)}{\sqrt f} \dd f
    \nonumber\\
    &=&  \frac{2\;\mathcal{C}(2\sqrt{f_N|t|})}{\sqrt{|t|}} ~~~,
    \nonumber
\end{eqnarray}
where $\mathcal{C}$ is the Fresnel cosine integral, $\mathcal{C} =\int_0^u \cos(\frac{\pi x^2}{2})\dd x$. Similarly the Fresnel sine integral is  $\mathcal{S} =\int_0^u \sin(\frac{\pi x^2}{2})\dd x$).

Applying the same process to the case $\Theta = 0 $ and $\Theta=\pi/2$ we find the discretized kernels to be:
\begin{subequations}
\begin{eqnarray}
    k_{0}(n\Delta t) &\approx&
    2\sqrt{\frac{2f_N}{Zh}} 
    \begin{cases}
        0 & \text{if } n = 0 \\
        \textrm{sgn}(n)\frac{\;\mathcal{S}(\sqrt{2|n|})}{\sqrt{ |n|}} & \text{if } n\neq0
    \end{cases}
    \nonumber
    \\
    k_{\pi/2}(n\Delta t) &\approx& 2\sqrt{\frac{2f_N}{Zh}} 
    \begin{cases}
        \sqrt{2} & \text{if } n = 0 \\
        \frac{\mathcal{C}(\sqrt{2|n|})}{\sqrt{ |n|}} & \text{if } n\neq0
    \end{cases}
    \nonumber
\end{eqnarray}    
\end{subequations}
with $n$ integer. We chose $\Theta=0$, i.e. we have worked with the kernel $k=k_0$.

\section{Quadratures measurement}
\label{Annexe:Quad_mesure}

All even moments of the time dependent quadrature operator 
\begin{equation} 
    \x_\Theta (t)=\frac{ \bdag(t) e^{i \Theta} + \b(t) e^{-i\Theta}}{\sqrt{2}}
\end{equation}
can be split in two sums (we kept an arbitrary phase $\Theta$ for the sake of generality). 
The first sum contains the non-symmetric (n.s.) terms, i.e. with a different number of creation and annihilation operators in all orders; the second contains the completely symmetric (c.s.) terms, i.e. with the same number of creation and annihilation operators in all orders:
\begin{eqnarray}
    \x^{2k}_{\Theta}(t)
    &=&
    \frac{1}{2^k}
    \sum_{l=0}^{k-1}
    \sum_{\text{n.s.}} \bigg[ \b(t)^{2k-l} \bdag[l](t) e^{i\Theta( -2k + 2l  )} + \hc \bigg]
\nonumber\\
    &+&
    \frac{1}{2^k}
    \sum_{\text{c.s.}} \b^{k}(t)\bdag[k](t)
    \;\;.
\label{eq:nsANDcs}
\end{eqnarray}
The c.s. can be expressed in terms of photon number operator only \cite{virally_discrete_2016,simoneau_photon_2017}:
\begin{equation}
\sum_{\text{c.s.}}=   
    \sum_{i=0}^k \frac{(2k)!}{(i!)^2(k-i)!2^{k-i}} \prod_{j=0}^{i-1}\big(\n(t)- j \big)
    \;\;.
\nonumber
\end{equation}
Since we are only interested in $k=1$ and $2$, we give their explicit form:
\begin{eqnarray}
    \x^{2}_{\Theta}(t)
    &=&
    \frac{\b^{2}(t) e^{-i2\Theta}+ \hc }{2}
    +
    \n(t) + \frac{1}{2}
\nonumber\\
    \x^{4}_{\Theta}(t)
    &=&
    \frac{\b^{4}(t) e^{-i4\Theta}+ \b^{3}(t) \bdag(t) e^{-i2\Theta} + \hc   }{4}
\nonumber\\
    &+&
    \frac{3}{2} \n^2(t) + \frac{3}{2}\n(t) + \frac{3}{4}
    \label{eq:TimeDependant_Quad}
    ~~~.~
\end{eqnarray}

Averaging incoherently, i.e. without any synchronisation between the generation of the radiation by the ac drive in the sample and the clock of the digitization, simplifies the expressions of the c.s. and n.s. terms.

If $\b$ describes a photon mode at frequency $f$, n.s. terms oscillate at frequencies $2(k-l)f$ and their time-average gives zero. This also holds for narrow-frequencies signals, so such terms were absent in \cite{virally_discrete_2016,simoneau_photon_2017}. This is no longer true for $\x_{\Theta}^4(t)$ for wideband signals, as for example $\b^3(t)\bdag(t)$ may contain a dc component provided the mode $\beta(f)$ is nonzero for some frequency $f$ as well as $3f$. In the particular case of bichromatic photons at frequencies $f_1$ and $f_2$, n.s. terms of $\x_{\Theta}^4(t)$ vanish upon time averaging except if $f_2=3f_1$. For general wideband signals, n.s. terms may contribute to the statistics of the quadratures. However for the shot noise we are considering here, n.s. terms are related to current correlators such as $\langle I(f_1)I(f_2)I(f_3)(-f_1-f_2-f_3)\rangle$ \cite{forgues_noise_2013,th_squeezing_inout} which are extremely small unless $f_1+f_2=f_p$ with $f_p$ the excitation frequency. In such a case $f_1+f_2+f_3>f_p$ is out of our detection bandwidth. Therefore n.s. terms never contribute here. In the absence of such terms, $\Theta$ becomes irrelevant, and we chose $\Theta=0$.

\clearpage
\section{Calibration}
\label{Annexe:Gain}

\begin{figure}[!ht]
	\centering
	\includegraphics[width=\linewidth]{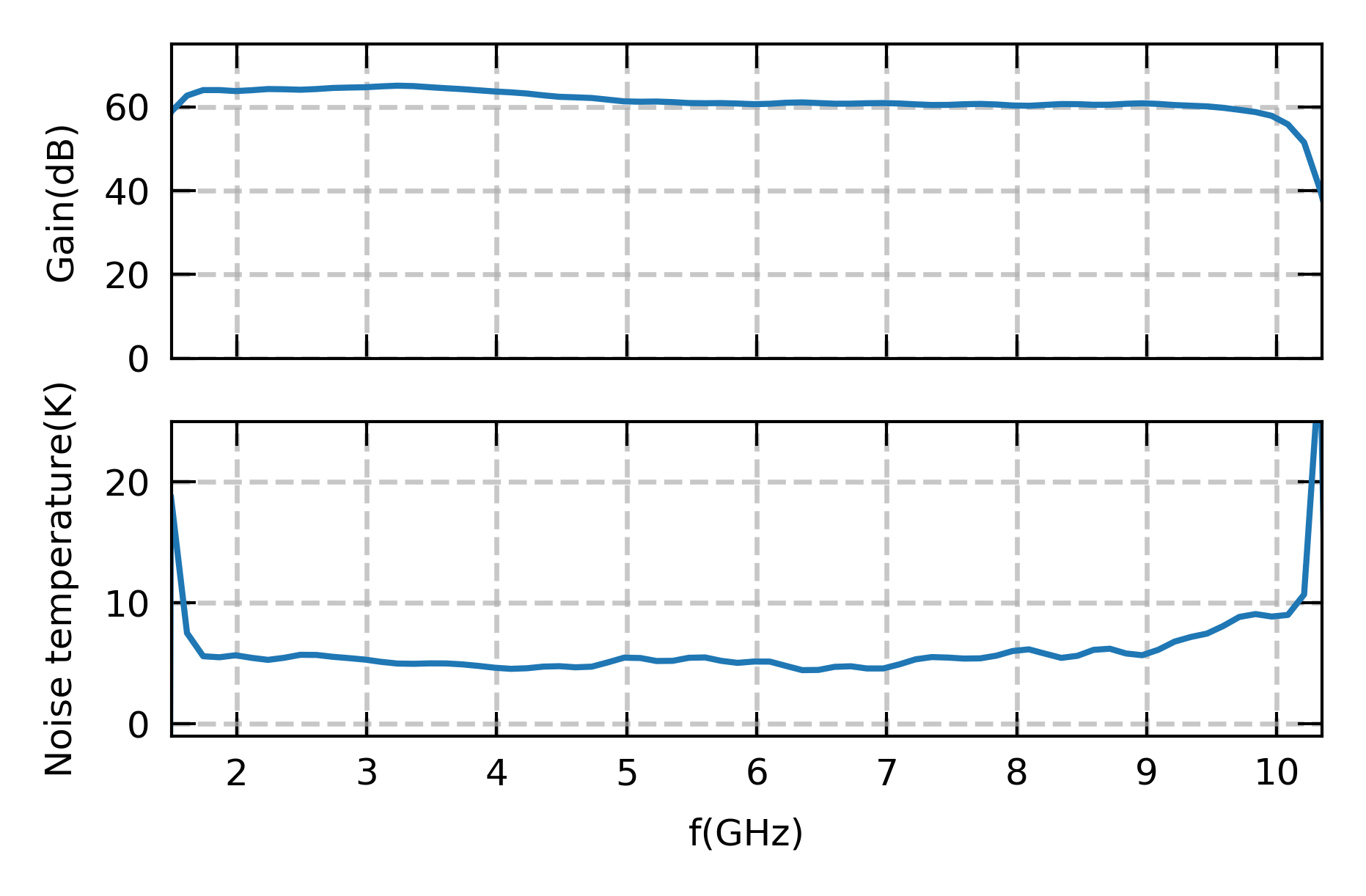}
	\caption{ 
        Frequency dependence of the properties of the amplification chain. Top: Power gain from the tunnel junction to the A/D converter.
        Bottom: Noise temperature.
    }
	\label{fig:Gain}
\end{figure}

The power gain and noise temperature $\Te$ are continuously measured and fitted from the full spectrum of current auto-correlations, 
as in \cite{Spietz_2003,forgues_non-classical_2016,simoneau_mesures_2021,esposito_TWPA2022}. Their frequency dependence is shown in Fig. \ref{fig:Gain}.
The slow drift in the noise temperature of the amplification chain is canceled out by interlacing experimental conditions with references ($I_{\text{dc}}=0$,$I_{\text{ac}}=0$).

\section{Nonlinearities}
\label{Annexe:Nonlin}
To prevent nonlinearity in the first amplification step, we ensure that the residual power from the pump is far below the 1dB compression point of the cryogenic amplifier, which is -39 dBm.
About 16 dB of the pump is reflected on the sample and is the attenuated by 17.5 dB by the 11 GHz low pass filter.
This means that of the -78 dBm of the pump reaching the 
sample, only -112 dBm is reaching the first amplifier, hence no nonlinearity is expected from that amplification step.
Moreover, the remaining 12 GHz is amplified by the first amplifier but falls out of the analog band of the rest of the amplification chain, so it has no influence on the measurement.

\begin{figure}[!ht]
	\centering
	\includegraphics[width=\linewidth]{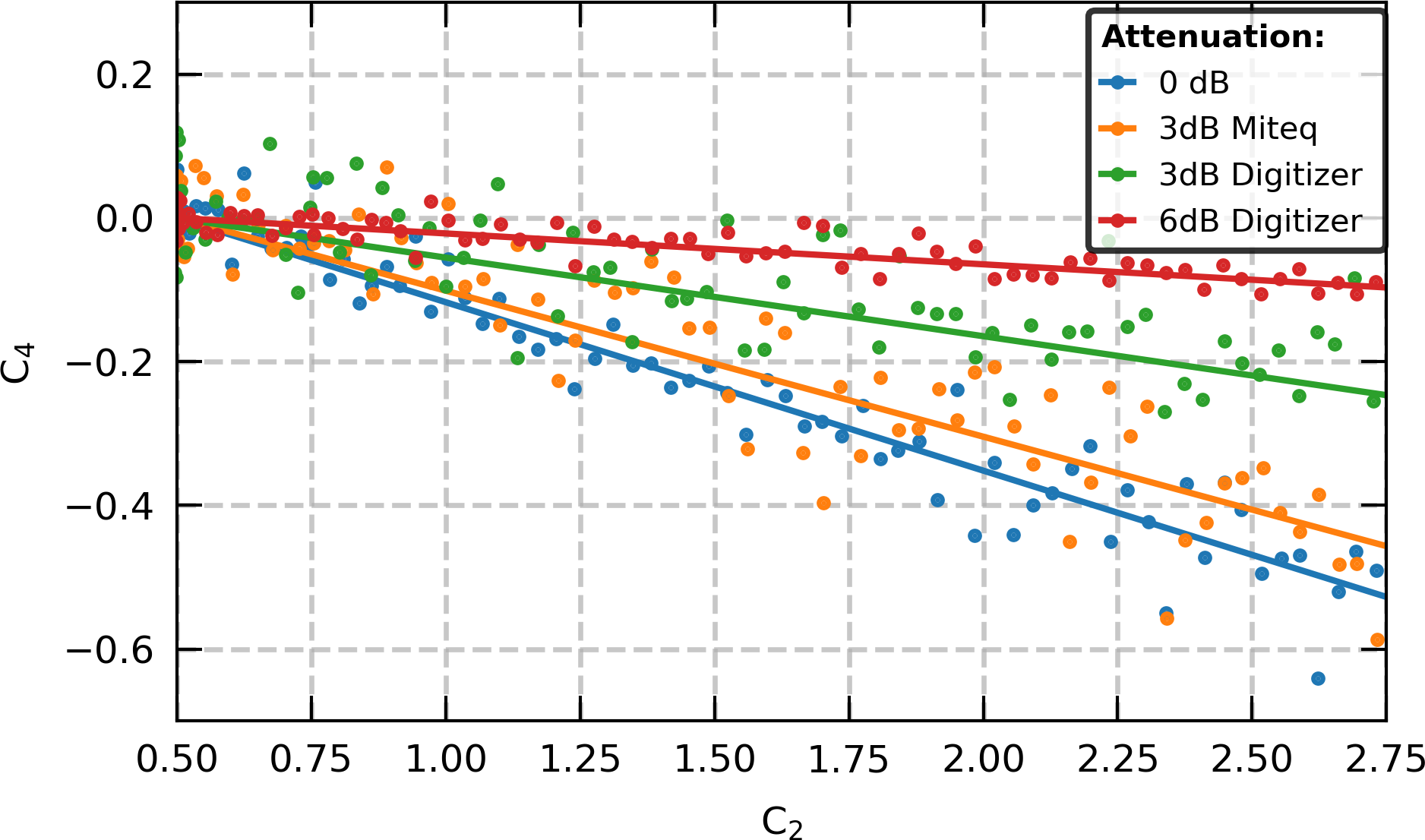}
	\caption{ Example of the effect of nonlinearities on the measurement of the 4th order cumulant as a function of the variance. The data presented is for a bichromatic 3\&9 GHz mode, but is representative of the nonlinearities measured in all cases.  
    $C_2$ and $C_4$ are cumulants of the unitless quadratures and hence are also unitless.}
	\label{fig:Nonlin}
\end{figure}

In an initial experiment we nevertheless observed nonlinearities affecting the measurement of the fourth order cumulant $C_4$. 
In Fig. \ref{fig:Nonlin}, we give an example of the effect of the nonlinearities on $C_4$ as a function of the variance $C_2$, for a bichromatic 3\&9 GHz mode. 
For vacuum fluctuations (leftmost part of the graph) and classical shot noise regime (rightmost part of the graph) $C_4$ should theoretically be extremely small.
We see that this was not the case.

The setup without any additional attenuation along the amplification chain is labeled '0 dB'.
If small nonlinearities are present in this regime, we expect to have $C_4 \propto (C_2+C_2^\textrm{amp})^3\propto C_2$, as observed. $C_2^\textrm{amp}$ is the noise of the amplifier, which is ranging from  $\sim100$ photons at 1 GHz to $\sim10$ photons at 10 GHz. 

Adding 3 dB of attenuation before the room temperature amplifier (1-12 GHz) and compensating with additional gain from the internal amplifier of the digitizer did not lead to any significant change in the observed nonlinearities('3 dB Miteq'). In contrast, reducing the gain of the digitizer ('3 dB Digitizer','6 dB Digitizer') clearly reduces the observed nonlinearities (without impacting the noise figure of the whole amplification chain). 
By reducing the incident power on the A/D converter by 4 we reduce the incident voltage by 2, essentially throwing away 1 bit out of the 10 bits of the digitizer. 
This sacrifice in dynamic range for linearity effectively resolved all nonlinearity related issues in measurements, rendering corrections unnecessary for the presented data.

\section{Electronic Temperature}
\label{Annexe:Te}
\begin{figure}[!ht]
	\centering
	\includegraphics[width=\linewidth]{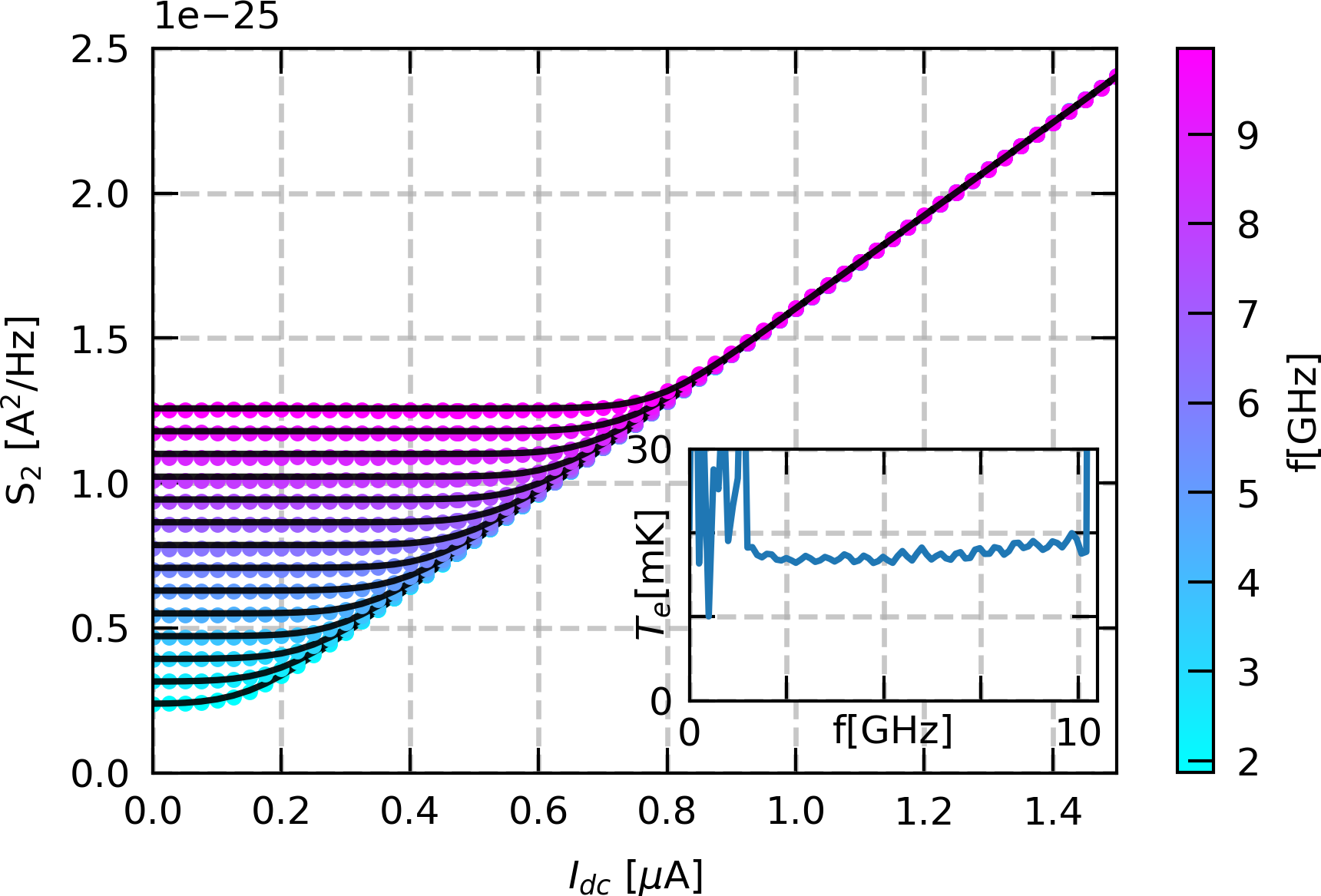}
	\caption{Noise spectral density $S_2(f)$, of our tunnel junction as a function of the dc bias current for frequencies between 2 and 10 GHz. Dots are measurements and lines are theoretical fits.
    Inset : The fitted electronic temperature as a function frequency.
    }
	\label{fig:SII_and_Te}
\end{figure}

We show  in Fig. \ref{fig:SII_and_Te} the sample noise spectral density for several frequencies. 
For each frequency, by fitting the noise vs. bias we can precisely determine the electron temperature $T_e$. We plot in the inset of Fig. \ref{fig:SII_and_Te} the obtained $T_e$ vs. frequency. It is almost frequency independent, and corresponds to $T_e=17.4$mK. 

\newpage
\section{Theoretical entanglement of formation for a vacuum squeezed state.}
\label{Annexe:Eof_sqvac}
The entanglement of formation grows at best logarithmically with the amount of squeezing for a perfect vacuum squeezed state .
\begin{figure}[!h]
    \centering
    \includegraphics[width= \linewidth]{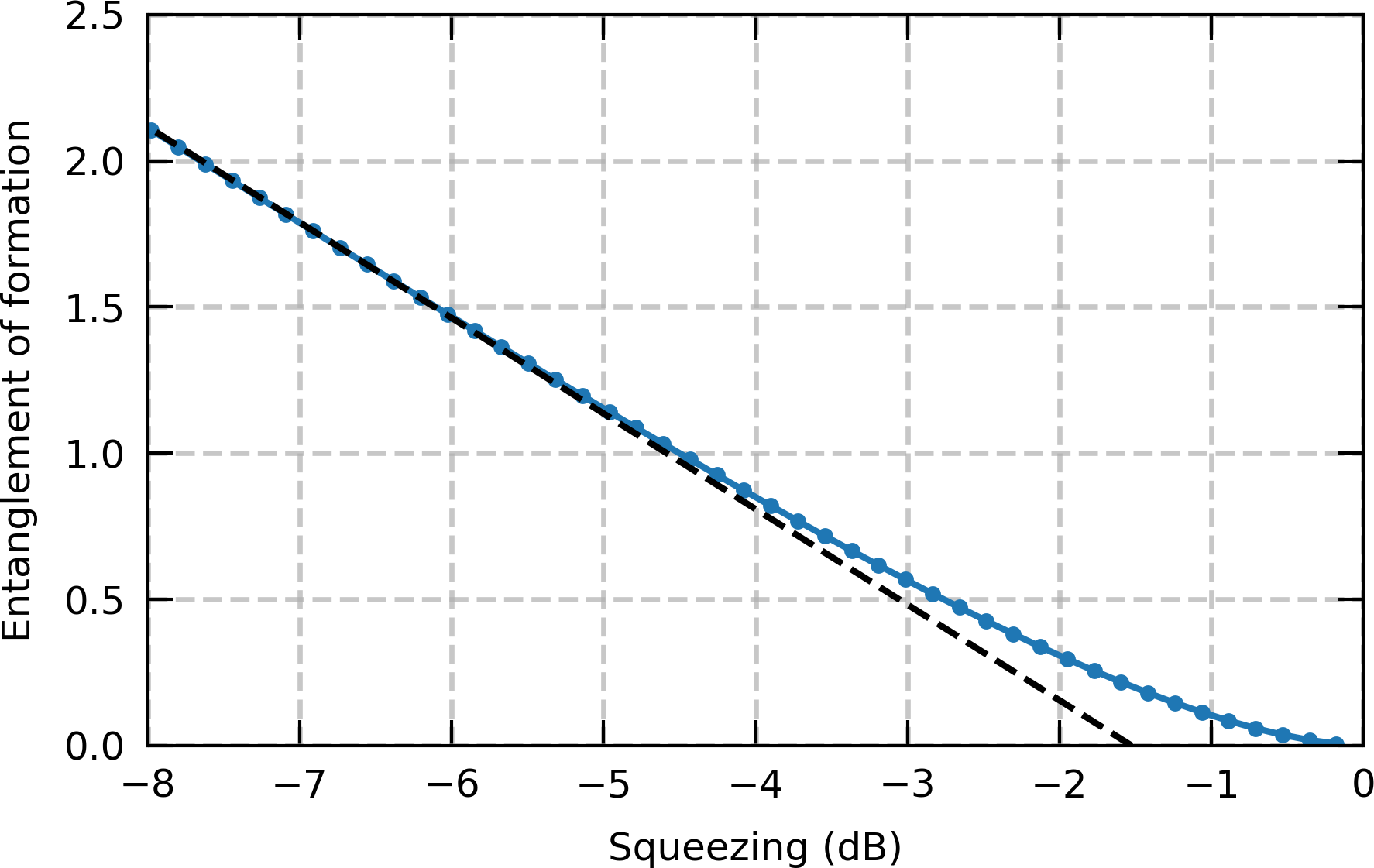}
    \caption{ Entanglement of formation for an ideal vacuum squeezed state. 
    The dashed line is a logarithmic fit at for high squeezing values.}
    \label{fig:Eof_sqvac}
\end{figure}

\section{Steering for the maximally squeezing conditions.}
\label{Annexe:Steering_13}

For completeness, here we show the results for steering in the same conditions as in Fig. \ref{fig:squeezing_spectrum}, i.e. when squeezing and entanglement of formation are maximal. 
\begin{figure}[!h]
    \centering
    \includegraphics[width= \linewidth]{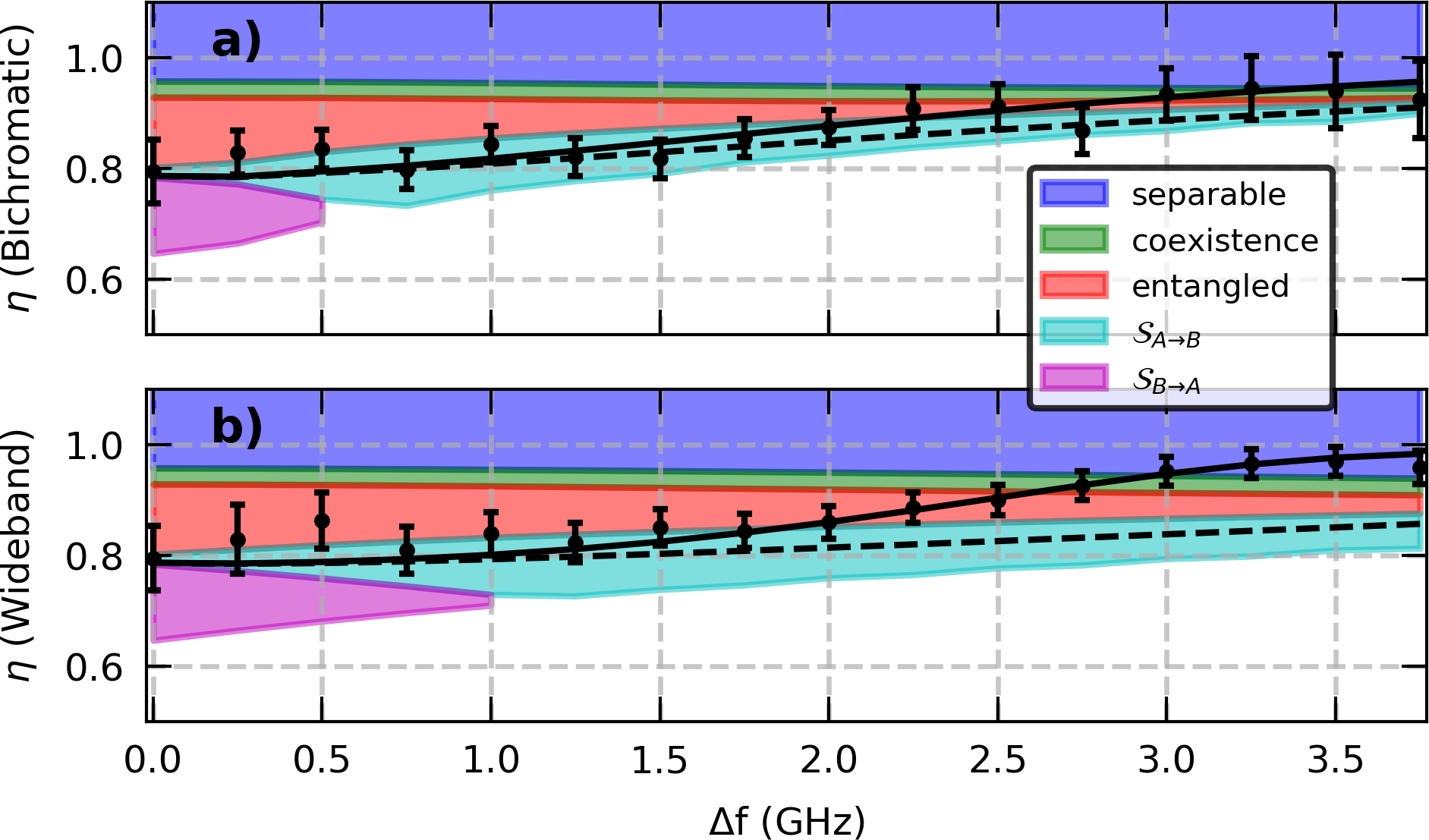}
    \caption{ 
    Classification of separability and steering of bipartite Gaussian states using the quantification based on marginals and global purities from \cite{Kogias_2015}.
    The measured $\eta$, '.' black markers,  is presented as function of the difference of frequencies from 6 GHz for bichromatic states (a) and wideband states (b). 
    Note that conditions optimal to observe steering are different from conditions of maximal squeezing. The tunnel junction is excited at frequency $f_p=12$ GHz with an RMS ac bias of $0.43~\mu$A and $\Vdc=hf_p/(2e)$.
    }
    \label{fig:Steering_13}
\end{figure}

\clearpage

\end{document}